\documentclass[a4paper,11pt]{article}
\usepackage[utf8]{inputenc}
\usepackage[T1,T2A]{fontenc}
\usepackage{amsmath,amsfonts,amssymb,mathtools}
\pdfoutput=1
\usepackage{graphicx}   
\usepackage{subcaption} 
\usepackage{float}
\usepackage{wrapfig}
\usepackage{hyperref}
\usepackage[normalem]{ulem}  

\usepackage{slashed}

\newcommand{\be}{\begin{equation}}
\newcommand{\ee}{\end{equation}}
\newcommand{\bea}{\begin{eqnarray}}
\newcommand{\eea}{\end{eqnarray}}
\newcommand{\bes}{\begin{subequations}}
\newcommand{\ees}{\end{subequations}}

\newcommand{\bc}{\begin{center}}
\newcommand{\ec}{\end{center}}

\begin{document}

\begin{center}
{\LARGE {\bf Type-II Seesaw Mechanism for Dirac Neutrinos and its Implications on $N_{\text{eff}}$ and Lepton Flavor Violation in a 3-3-1 model}}
\end{center}

\vspace{1cm}
\begin{center}
{\bf Vinícius Oliveira$^{\,1\,,2\,,3}$, Patricio Escalona$^{\,4}$, Lucia Angel$^{\,5\,,6\,,7}$, C. A. de S. Pires$^{\,4}$, Farinaldo S. Queiroz$^{5, 6 ,7 ,8 }$}
\end{center}

\begin{center}
  \vspace*{0.25cm}
  $^1$ \it{Departamento de Física da Universidade de Aveiro and \\
  Center for Research and Development in Mathematics and Applications (CIDMA),\\
  Campus de Santiago, 3810-183 Aveiro, Portugal}\\
  \vspace{0.2cm}
  $^2$ \it{Department of Physics, Lund University, 221 00 Lund, Sweden}\\
  \vspace{0.2cm}
  $^3$ \it{Laboratório de Instrumentação e Física Experimental de Partículas (LIP), \\
  Universidade do Minho, 4710-057 Braga, Portugal}\\
  \vspace{0.2cm}
 $^4$ \it{ Departamento de Física, Universidade Federal da Paraíba, \\
  Caixa Postal 5008, 58051-970, João Pessoa, PB, Brazil,}\\
  \vspace{0.2cm}
  $^5$ \it{International Institute of Physics, Universidade Federal do Rio Grande do Norte,\\
Campus Universitário, Lagoa Nova, Natal-RN 59078-970, Brazil}\\
  \vspace{0.2cm}
  $^6$ \it{Millennium Institute for Subatomic Physics at the High-Energy Frontier (SAPHIR) of ANID, Fernández Concha 700, Santiago, Chile}\\
  \vspace{0.2cm}
  $^7$ \it{Departamento de Física, Universidade Federal do Rio Grande do Norte \\
  59078-970, Natal, RN,  Brazil}\\
  \vspace{0.2cm}
  $^8$ \it{Universidad de La Serena, \\
  Casilla 554, La Serena, Chile} \\
  \vspace{0.2cm}
\end{center}

\vspace{1.5cm}

\begin{center} {\bf Abstract} \end{center}
\noindent

In this study, we implement the type-II seesaw mechanism for Dirac neutrino masses within the framework of a 3-3-1 model. To this end, we introduce a scalar sextet and impose both lepton number conservation and invariance under a discrete $Z_2$ symmetry in the Lagrangian.  This mechanism naturally generates small Dirac neutrino masses by allowing the soft breaking of the $Z_2$ symmetry through a unique term in the scalar potential, while preserving lepton number. Specifically, we explore the realization of this model at low-energy scales. Regarding flavor implications, we analyze its contributions to the rare decay processes $\mu \to e \gamma$ and $\mu \to \bar e ee$. In the cosmological context, we analyze the influence of right-handed neutrinos on the effective number of neutrino species, $N_\text{eff}$, through interactions mediated by the $Z^{\prime}$ boson. Our findings establish a lower bound of $m_{Z^{\prime}} > 4.4$ TeV, which slightly exceeds the current lower limit set by the Large Hadron Collider (LHC).
\vspace{1cm}

\newpage 
\tableofcontents

\section{Introduction}
\label{sec:Introduction}
Neutrino oscillation experiments have definitively shown that neutrinos are massive particles \cite{Kajita:2016cak} with masses in the sub-eV scale \cite{ParticleDataGroup:2024cfk}. These small masses, in turn, call for an explanation. Moreover, the longstanding problem of the neutrinos nature, whether they are Dirac or Majorana particles, persists to this day. Due to the lack of experimental evidence regarding the intrinsic nature of neutrinos, advancements in the field have largely been driven by theoretical assumptions about their fundamental properties. Following this approach, the prevailing hypothesis is that neutrinos are Majorana particles. This is motivated by the fact that dimension-five effective operators generating Majorana masses naturally arise within the Standard Model (SM) while respecting gauge symmetries \cite{Weinberg:1979sa}. 

A key experimental signature of Majorana neutrinos is neutrinoless double beta decay \cite{Simkovic:2013kna}, a process absent in the case of Dirac neutrinos. Despite extensive searches for detecting the neutrinoless double beta decay process conducted over several decades, the absence of definitive positive results \cite{Cirigliano:2022oqy} has prompted the consideration of alternative scenarios, such as seesaw-like mechanisms for generating Dirac neutrino masses

The seesaw mechanism is the most widely studied framework for generating Majorana neutrino masses at tree-level. There are several types of seesaw mechanisms, for instance: type-I \cite{Minkowski:1977sc,Mohapatra:1979ia}, type-II \cite{Ma:1998dx} and inverse seesaw \cite{Mohapatra:1986bd} mechanisms. However, the construction of tree-level seesaw mechanisms for Dirac neutrino masses has received relatively little attention \cite{CentellesChulia:2016rms,Bonilla:2017ekt,Borah:2017dmk,Borah:2022obi}. This limited interest stems from two main reasons: first, building viable Dirac seesaw models is not as straightforward as for Majorana ones; second, as mentioned above, Dirac neutrinos lack a distinctive experimental signature.

In this work we address these issues within a particular variant of the 3-3-1 model. The so-called 3-3-1 models are gauge extensions of the SM based on the  $\text{SU}(3)_C \times \text{SU}(3)_L \times \text{U}(1)_N$ gauge group. Several versions of 3-3-1 models have been proposed, but the most well-developed ones are the minimal 3-3-1  model \cite{Frampton:1992wt,Pisano:1992bxx} and the 3-3-1 model with right-handed neutrinos (331RHN) \cite{Singer:1980sw,Montero:1992jk,Foot:1994ym}. A key motivation for studying these models is their satisfactory explanation of the number of generations in the SM \cite{Pisano:1996ht}. Those models also have clear predictions for flavor-changing neutral current (FCNC) processes involving quarks \cite{Buras:2012dp,Cogollo:2012ek,Ferreira:2019qpf,Queiroz:2016gif,Arcadi:2017xbo,CarcamoHernandez:2022fvl,deJesus:2023lvn}.  Another notable feature of these models is that each 3-3-1 model has three variants, distinguished by which of the three families of quarks transforms as a triplet by the SU$(3)_L$ symmetry to ensure anomaly cancellation \cite{Long:1999ij,Oliveira:2022vjo}.  Moreover, the minimal 3-3-1 model and the 331RHN  explain electric charge quantization \cite{deSousaPires:1998jc,deSousaPires:1999ca}. All these motivations make the 3-3-1 models, and especially the 331RHN, a compelling framework for exploring physics beyond the SM.

In the original version of the 331RHN, neutrinos are massless particles \cite{Singer:1980sw,Montero:1992jk,Foot:1994ym}. However,  since the discovery of neutrino masses via neutrino oscillation experiments, various extensions of the 331RHN have been proposed to account for neutrino masses. In the majority of these proposals, neutrinos are considered Majorana particles \cite{Dias:2005yh,Cogollo:2008zc,Cogollo:2009yi,Hepburn:2022pin,Dias:2012xp,CarcamoHernandez:2013krw,Boucenna:2014ela,deSousaPires:2018fnl,CarcamoHernandez:2020pnh,Doff:2024ovy}. Dirac neutrino masses induced by the seesaw mechanism within 3-3-1 models have been addressed only in  two works \cite{Reig:2016ewy,Valle:2016kyz}. 
 
In this work, we propose an extension of the 331RHN to generate Dirac neutrino masses, through a mechanism analogous to the type-II seesaw mechanism for Majorana neutrinos. This is achieved by introducing a scalar sextet to the original scalar content of the 331RHN and imposing that the model’s Lagrangian  preserves the total lepton number and remains invariant under a discrete $Z_2$ symmetry with specific field transformations. We focus in a low-scale seesaw mechanism with the sextet of scalars at the TeV scale. As phenomenological implications, we obtain the mechanism’s prediction for the branching ratios of the rare lepton decay processes  $\mu \to e\gamma$, $\mu \to \bar e ee$  and derive a constraint on the mass of the new neutral gauge boson $Z^{\prime}$ based on the contribution of right-handed neutrinos and the modifications of left-handed neutrino interactions to the effective number of neutrino species, $N_\text{eff}$.

The work is organized as follows: In \autoref{sec:general_aspects} we present general aspects of the model.  In \autoref{sec:seesaw_mechanism} we develop the seesaw mechanism. In \autoref{sec:LFV} we study the rare decay processes $\mu \to e\gamma$ and  $\mu \to \bar e ee$ mediated by charged scalars of the model.  In \autoref{sec:Neff}  we compute the effects of a light Dirac neutrino with specific new interactions on $N_\text{eff}$ and derive bounds on the energy scale at which the 3-3-1 symmetry spontaneously breaks down to the SM symmetry. Finally, in \autoref{sec:conclusion}, we summarize our conclusions. 

\section{General aspects of the model}
\label{sec:general_aspects}
The leptonic content of the model is arranged into lepton triplets and singlets in the following form \cite{Singer:1980sw,Montero:1992jk,Foot:1994ym}:
\begin{equation}
f_{aL}= \begin{pmatrix}
\nu_{a}     \\
l_{a}       \\
\nu^{c}_{a} \\
\end{pmatrix}_{L} \sim (1,3,-1/3), \quad e_{aR}\sim (1,1,-1),
\label{leptons}
\end{equation}
with $a=1,2,3$ representing the three SM generations of leptons.

In the hadronic sector, anomaly cancellation requires one quark family to transform differently from the other two. This condition allows for three possible arrangements of the quark families \cite{Oliveira:2022vjo}. Here, we consider the arrangement in which the third generation transforms as a triplet of SU$(3)_L$, while the other two transforms as anti-triplet, 
\begin{eqnarray}
&&Q_{iL} = \left (
\begin{array}{c}
d_{i} \\
-u_{i} \\
d^{\prime}_{i}
\end{array}
\right )_L\sim(3\,,\,\bar{3}\,,\,0)\,,u_{iR}\,\sim(3,1,2/3),\,\,\,\nonumber \\
&&\,\,d_{iR}\,\sim(3,1,-1/3)\,,\,\,\,\, d^{\prime}_{iR}\,\sim(3,1,-1/3),\nonumber \\
&&Q_{3L} = \left (
\begin{array}{c}
u_{3} \\
d_{3} \\
u^{\prime}_{3}
\end{array}
\right )_L\sim(3\,,\,3\,,\,1/3),u_{3R}\,\sim(3,1,2/3),\nonumber \\
&&\,\,d_{3R}\,\sim(3,1,-1/3)\,,\,u^{\prime}_{3R}\,\sim(3,1,2/3),
\label{quarks} 
\end{eqnarray}
where  the index $i=1,2$ is restricted to the first two generations. The primed quarks are new heavy quarks with the usual electric charges $(+\frac{2}{3}, -\frac{1}{3})$. 

The original  scalar sector of the 331RHNs is composed of three triplets,
\begin{eqnarray}
&&\eta = \left (
\begin{array}{c}
\eta^0 \\
\eta^- \\
\eta^{\prime 0}
\end{array}
\right ),\,\rho = \left (
\begin{array}{c}
\rho^+ \\
\rho^0 \\
\rho^{\prime +}
\end{array}
\right ),\,
\chi = \left (
\begin{array}{c}
\chi^0 \\
\chi^{-} \\
\chi^{\prime 0}
\end{array}
\right ),
\label{fermiontriplets} 
\end{eqnarray}
with $\eta$ and $\chi$ transforming as $(1\,,\,3\,,\,-1/3)$, and $\rho$ as $(1\,,\,3\,,\,2/3)$.

The simplest version of the model is achieved by requiring the Lagrangian to be symmetric under a $Z_2$ discrete symmetry, with the fields transforming as
\begin{equation}
(\rho\,,\,\chi\,,\,e_{a_R}, u_{a_R}, Q_{3_L}\,,\, d^{\prime}_{i_R}) \to -(\rho\,,\,\chi\,,\,e_{a_R}, u_{a_R}, Q_{3_L}\,,\, d^{\prime}_{i_R})\,,
\label{Z2}
\end{equation}
where $a=1,2,3$. This symmetry provides the most efficient configuration of Yukawa interactions necessary to generate masses for all charged fermions in the model,
\begin{eqnarray}
-{\cal L}^Y &=&f_{ij} \bar Q_{i_L}\chi^* d^{\prime}_{j_R} +f_{33} \bar Q_{3_L}\chi u^{\prime}_{3_R} + g_{ia}\bar Q_{i_L}\eta^* d_{a_R} +h_{3a} \bar Q_{3_L}\eta u_{a_R}\nonumber \\
&& +g_{3a}\bar Q_{3_L}\rho d_{a_R}+h_{ia}\bar Q_{i_L}\rho^* u_{a_R}+ G_{l}\bar f_{l_L} \rho e_{l_R}  +  \mbox{H.c} \,.
\label{yukawa1}
\end{eqnarray}
For simplicity, we consider the charged leptons in a diagonal basis.

The most general potential that preserves lepton number and is invariant under $Z_2$ consists of the following set of terms:
\begin{eqnarray} 
V(\eta,\rho,\chi)&=&\mu_\chi \chi^{\dagger} \chi +\mu_\eta \eta^{\dagger} \eta
+\mu_\rho \rho^{\dagger} \rho+\lambda_1(\chi^{\dagger}\chi)^2 +\lambda_2(\eta^{\dagger}\eta)^2
+\lambda_3(\rho^{\dagger}\rho)^2 \nonumber \\
&&+ \lambda_4(\chi^{\dagger}\chi)(\eta^{\dagger}\eta)
+\lambda_5(\chi^{\dagger}\chi)(\rho^{\dagger}\rho)+\lambda_6
(\eta^{\dagger}\eta)(\rho^{\dagger}\rho) \nonumber\\
&&+\lambda_7(\chi^{\dagger}\eta)(\eta^{\dagger}\chi)
+\lambda_8(\chi^{\dagger}\rho)(\rho^{\dagger}\chi)+\lambda_9
(\eta^{\dagger}\rho)(\rho^{\dagger}\eta)   \nonumber\\
&&-\left( \dfrac{f}{\sqrt{2}} \epsilon^{ijk} \eta_i \rho_j \chi_k +  \mathrm{H.c.}\right). 
\label{PPLN}
\end{eqnarray}

The Higgs spectrum corresponding to this potential is examined in Ref.~\cite{Pinheiro:2022bcs} for the case where only $\eta^0$, $\rho^0$, and $\chi^{\prime 0}$ acquire non-zero vacuum expectation values (VEVs), denoted as $v_\eta$, $v_\rho$, and $v_{\chi^{\prime}}$, respectively. After the spontaneous symmetry breaking (SSB) of the $\text{SU}(3)_L \times \text{U}(1)_N \to \text{U}(1)_\text{QED}$, the scalar potential yields $8$ Goldstones which are eaten by the gauge bosons, namely, the CP-odd part of $\chi^{\prime 0}$ is eaten by $Z^{\prime}$, a linear combination of the CP-odd part of $\eta^0$ and $\rho^0$ is eaten by $Z$, a combination of $\eta^{\pm}$ and $\rho^{\pm}$ is eaten by $W^{\pm}$, a combination of $\rho^{\prime \pm}$ and $\chi^{\pm}$ is eaten by $W^{\prime \pm}$, and finally, a combination of $\eta^{\prime 0}$ and $\chi^0$ is eaten by $U^{0}$ and $ U^{0 \dagger}$.

Concerning the gauge boson spectrum of the electroweak sector, the $\text{SU}(3)_L \times \text{U}(1)_N$ symmetry group implies the existence of nine such particles: namely, eight associated with associated to $\text{SU}(3)_L$, denoted  $W^i_\mu$ ($i=1,..,8$),  and one associated to U$(1)_N$, called $B_\mu$. After the SSB, the gauge bosons $W^{1}_\mu$  and $W^2_\mu$ mix to form the standard charged $W^{\pm}_\mu$, $W^{6}_\mu$  and $W^7_\mu$ mix to form the new charged $W^{\prime \pm}_\mu$,  $W^{4}_\mu$  and $W^5_\mu$ mix to form the non-Hermitian $U^0_\mu$ and $U^{0 \dagger}_\mu$, and finally, $W^3_\mu$, $W^8_\mu$ and $B_\mu$, mix to form the neutral gauge bosons $Z$, $Z^{ \prime}$, and the photon $A_\mu$ \cite{Long:1995ctv,Long:1995rkv}
\begin{eqnarray}
    &&A_\mu= s_W W^3_\mu +c_W\left(-\frac{t_W}{\sqrt{3}}W^8_\mu + \sqrt{1-\frac{t_W^2}{3}}B_\mu\right),\nonumber \\
   && Z_\mu= c_W W^3_\mu +s_W\left(-\frac{t_W}{\sqrt{3}}W^8_\mu + \sqrt{1-\frac{t_W^2}{3}}B_\mu\right),\nonumber \\
   &&Z^{\prime}_\mu = \sqrt{1-\frac{t_W^2}{3}}W^8_\mu +\frac{t_W}{\sqrt{3}}B_\mu,
\end{eqnarray}
where $s_W=\sin\theta_W$, $c_W=\cos\theta_W$ and  $t_W=\tan\theta_W$, with $\theta_W$ being the Weinberg angle. The gauge bosons masses are given by
\begin{eqnarray}
&&m^2_W=\frac{g^2}{4}(v^2_\eta + v^2_\rho)\,\,\,,\,\,\, m^2_{W^{\prime}}=\frac{g^2}{4}(v^2_\eta + v^2_{\chi^{\prime}})\,\,\,,\,\,\, m^2_{U^0}=\frac{g^2}{4}(v^2_\rho + v^2_{\chi^{\prime}})\nonumber \\
&&m^2_Z=\frac{g^2}{4c^2_W}(v^2_\eta + v^2_\rho)\,\,\,,\,\,\,m^2_{Z^{\prime}}=\frac{g^2}{3-4s^2_W}v^2_{\chi^{\prime}},
\end{eqnarray}
where  $v^2_\eta +v^2_\rho=v^2_\text{ew}$, with $v_\text{ew}=246$ GeV. Concerning the gauge couplings, the coupling constant of SU$(3)_L$ is  $g$ and the coupling constant of U$(1)_N$ is $g_N$. After the SSB $\text{SU}(3)_L \times \text{U}(1)_N \to \text{SU}(2)_L \times U(1)_Y$, the gauge coupling of $U(1)_Y$, denoted by $g^{\prime}$, is related to $g_N$ by  $\frac{1}{g^{\prime 2}}=\frac{1}{3g^2} + \frac{6}{g^2_N}$. Details can be found in Ref. \cite{Long:1995ctv}

Whenever $v_{\chi^{\prime}} \gg v_\eta, v_\rho$ we have $m_{W^{\prime}} \approx m_{U^0} \approx \frac{\sqrt{3-4s^2_W}}{2}m_{Z^{\prime}}\approx 0.72 m_{Z^{\prime}}$. This implies that any constraint on $m_{Z^{\prime}}$ indirectly applies to $m_{W^{\prime}}$ and $m_{U^0}$ as well. Current collider constraint  demands $m_{Z^{\prime}} \geq 4$ TeV. However the more stringent constraints on this model  arise from FCNC processes in the hadronic sector. This is because, in 3-3-1 models, FCNCs are an unavoidable consequence of anomaly cancellation (see \cite{Oliveira:2022vjo, Escalona:2025rxu}).

\section{The seesaw mechanism and neutrino masses}
\label{sec:seesaw_mechanism}
Despite of the complexity of the model, the original version predicts massless neutrinos. To generate neutrino masses we add a sextet of scalars \cite{Montero:1992jk} as follows:
\begin{eqnarray}
S=\frac{1}{\sqrt{2}}\left(\begin{array}{ccc}
\sqrt{2}\, \Delta^{0} & \Delta^{-} & \Phi^{0} \\
\newline \\
\Delta^{-} & \sqrt{2}\, \Delta^{--} & \Phi^{-} \\
\newline \\
\Phi^{0} & \Phi^{-} & \sqrt{2}\, \sigma^{0} \end{array}\right)\,,
\label{Scalarsextet}
\end{eqnarray}
with $S$ transforming as $(1,6,-2/3)$.

Imposing lepton number conservation on the  first two terms of \autoref{C-N-currents} implies that the gauge bosons $W^{\prime \pm}$ and $U^0$ must carry two units of lepton number each, namely $L(W^{\prime \pm})= L(U^0)=-2$. The charged and neutral currents involving standard quarks, new quarks and the new gauge bosons $W^{\prime \pm}$ and $U^0$,namely $\frac{g}{\sqrt{2}} \bar q_L \gamma^\mu q^{\prime}_L W^{\prime +}$ and $\frac{g}{\sqrt{2}}\bar q_L \gamma^\mu q^{\prime}_L U^0_\mu$ \cite{Long:1995ctv},  require that $d^{\prime}_i$ and $u^{\prime}_3$ also carry two units of lepton number. Extending lepton number conservation to the Yukawa sector leads to the assignment of two units of lepton number to the scalars $\eta^{\prime0}, \chi^0 , \Delta ^0 $ and $\sigma^0$. We refer to these particles as bileptons. 

In view of this, the most general potential that preserves lepton number and is invariant under $Z_2$ is
\begin{eqnarray} 
V(\eta,\rho,\chi, S)&=& V(\eta,\rho,\chi) +\mu_S^2 Tr[S^\dagger S]+ \lambda_{10} Tr[S^\dagger S]^2 + \lambda_{11} Tr[S^\dagger SS^\dagger S] \nonumber\\
&&+ (\lambda_{12}\eta^\dagger\eta +\lambda_{13}\rho^\dagger\rho +\lambda_{14}\chi^\dagger\chi  )Tr[S^\dagger S]  \nonumber\\
&&+\lambda_{15} \eta^\dagger S S^\dagger \eta+ \lambda_{16} \rho^\dagger S S^\dagger \rho+\lambda_{17} \chi^\dagger S S^\dagger \chi .
\end{eqnarray}

We emphasize that, with this set of scalars, we can write an unique soft breaking term in the potential that preserves lepton number but explicitly violates $Z_2$, namely:
\begin{equation}
    -\dfrac{M}{\sqrt{2}}\eta^T S^{\dagger} \chi + \mathrm{H.c.} \,.
    \label{PVZ2}
\end{equation}
This soft breaking term plays a central role in realizing the seesaw mechanism under consideration. The scalar spectrum of the model is detailed in \autoref{appendix:A1}.

To prevent spontaneous breaking of lepton number at all orders of perturbation theory, we require that only $\Phi^0$ of the sextet $S$ acquires non-zero VEV. Consequently,  the set of  scalars that develop non-zero VEVs are $\eta^0\,\,, \rho^0\,\,, \chi^{\prime 0}$, and $\Phi^0$. These fields are redefined by shifting  in the usual way,
\begin{equation}
  \eta^0, \rho^0, \chi^{\prime 0}, \Phi^0 =\frac{1}{\sqrt{2}}\left(v_{\eta, \rho,\chi^{\prime},\Phi^0\,\,}+R_{\eta, \rho,\chi^{\prime},\Phi^0}+iI_{\eta, \rho,\chi^{\prime},\Phi^0 }\right).
\end{equation}

In light of the above considerations, the scalar potential yields the following set of  equations, which ensure that it develops a stable  minimum:
\begin{eqnarray}
&&  \mu _{{\chi} }{}^2+\lambda_1 v_{{\chi^\prime} }^2+\frac{1}{2} \lambda _4 v_{\eta }^2 +\frac{1}{2} \lambda _5 v_{\rho }^2 +\frac{1}{2} \lambda _{14}v_{\Phi }^2+\frac{1}{4} \lambda _{17} v_{\Phi }^2 -\frac{f v_{\eta } v_{\rho }}{2v_{{\chi^\prime} }}-\frac{1}{2\sqrt{2}}\frac{ M v_{\eta } v_{\Phi }}{v_{{\chi^\prime} }}=0,\nonumber \\
   && \mu _{\eta }{}^2+\lambda _2 v_{\eta }^2+\frac{1}{2} \lambda _4 v_{{\chi^\prime} }^2+\frac{1}{2} \lambda _6  v_{\rho }^2+\frac{1}{2} \lambda _{12} v_{\Phi
   }^2+\frac{1}{4} \lambda _{15} v_{\Phi }^2-\frac{f v_{\rho } v_{{\chi^\prime} }}{2 v_\eta}-\frac{1}{2\sqrt{2}} \frac{M v_{{\chi^\prime} } v_{\Phi }}{v_\eta}=0, \nonumber \\
   && \mu _{\rho }{}^2+\lambda _3 v_{\rho }^2+\frac{1}{2} \lambda _5 v_{{\chi^\prime} }^2+\frac{1}{2} \lambda _6 v_{\eta }^2 +\frac{1}{2} \lambda _{13} v_{\Phi }^2-\frac{f v_{\eta } v_{{\chi^\prime} }}{2v_\rho}=0, \label{eq:mc}   \\ \nonumber
   && \mu _s{}^2+(\lambda _{10} +\frac{ \lambda _{11}}{2}) v_{\Phi }^2+\frac{ \lambda _{12}}{2} v_{\eta }^2 +\frac{ \lambda _{13}}{2} v_{\rho }^2+\frac{ \lambda _{14}}{2} v_{{\chi^\prime} }^2 + \frac{ \lambda _{15}}{4} v_{\eta}^2 +\frac{ \lambda _{17}}{4} v_{{\chi^\prime} }^2 -\frac{1}{2\sqrt{2}}\frac{M v_{\eta } v_{{\chi^\prime} }}{v_\Phi} = 0.
\end{eqnarray}

The Yukawa interaction is extended by adding the term
\begin{eqnarray}
-{\cal L}^Y \supset G^\nu _{ab} \bar f_{a_L}S^*\,(f_{b_L})^c  +  \mbox{H.c.} \,,
\label{yukawa2}
\end{eqnarray}
where $a,b=1,2,3$. Such Yukawa interaction leads to following  Dirac mass term:
\begin{equation}
    {\cal L}^{\text{Dirac}}_\nu = G^\nu_{ij}  \bar \nu_{i_L} \nu_{j_R} v_\Phi + \mathrm{H.c.} \, .
    \label{nuDmass1}
\end{equation}
The Dirac neutrino mass above is proportional to $v_\Phi$. Dirac neutrino masses are generally regarded as less favorable than Majorana masses, since the latter can more naturally attain small values via  seesaw mechanism, either through suppression of mass scales or VEVs. In what follows we show that the 331RHN model extended with the scalar sextet, supports a natural small $v_\Phi$, in analogy with the  canonical type-II seesaw mechanism, now applied to the Dirac neutrino case.

The parameter $M$ in  \autoref{PVZ2} represents the energy scale in which the $Z_2$ symmetry is explicitly broken. Under the assumption that  $M \ll v_\eta\,,\,v_\rho$, the final terms in the first and second relations in \autoref{eq:mc} become negligible. Conversely, in the fourth relation, the first and final terms dominate, resulting in
\begin{equation}
    v_\Phi \approx M\frac{ v_{\eta } v_{{\chi^\prime} }}{2\sqrt{2} \mu^2_S}.
    \label{seesaw}
\end{equation}
This expression is analogous  to that of the canonical type-II seesaw mechanism case \cite{Ma:1998dx}. 

In the canonical type-II seesaw mechanism, if the  energy scale associated with the explicit breaking of the global symmetry lies in sub-keV and the masses of the particles driving the mechanism are around the TeV scale,  an intriguing coincide emerges. The interplay between these two scales naturally leads to a VEV at the eV scale, which is precisely the order of magnitude required to account for neutrino masses. Following this reasoning to our case, we find that a small $v_\Phi \sim $ eV arises naturally from \autoref{seesaw}, given that the typical 3-3-1 energy scale lies around the TeV regime. For illustration,  consider benchmark values   $\mu_S= v_{{\chi^\prime} }\approx 10$ TeV  and $v_\eta \approx 10^2$ GeV. Under these assumptions, $v_\Phi\sim $ eV is obtained for $M=0.1$ keV. 

In this way, we have successfully adapted the canonical type-II seesaw mechanism, which suppresses VEVs associated with the explicit violation of global symmetries, to a scenario involving the explicit breaking of discrete symmetry within the 331RHN framework. This adaptation is significant as it extends the applicability of the mechanism beyond the generation of Majorana neutrino masses. Specifically, within the 331RHN, the resulting suppressed VEV naturally gives rise to small Dirac neutrino masses.

We now return to the Dirac mass term introduced in \autoref{nuDmass1} and express it in terms of the corresponding mass matrix:
\begin{equation}
    {\cal L}^{\text{Dirac}}_\nu = \bar \nu_L M_\nu \nu_R  + \mathrm{H.c.} \,\,\,\,\,\,\mbox{with} \,\,\,\,\, M_\nu = G^\nu v_\Phi \,,
\end{equation}
where $M_\nu$ is the neutrino mass matrix written in the flavor basis,  
\begin{eqnarray}
&&\nu_L = \left (
\begin{array}{c}
\nu_{e_L} \\
\nu_{\mu_L} \\
\nu_{\tau_L}
\end{array}
\right )\,\,\,\,\,,\,\,\,\nu_R = \left (
\begin{array}{c}
\nu_{e_R} \\
\nu_{\mu_R} \\
\nu_{\tau_R}
\end{array}
\right ).\,
\label{fermionstriplets} 
\end{eqnarray}
By performing the following change of basis,
\begin{equation}
    n_L=U^{\nu \dagger}_L \nu_L\,\,\,\,\,,\,\,\,\,\, n_R=U^{\nu \dagger}_R \nu_R,
    \label{CB}
\end{equation}
we obtain
\begin{equation}
    {\cal L}^{\text{Dirac}}_n = \bar n_L M^D_\nu n_R  + \mathrm{H.c.} \,\,\,\,\,\, \mbox{with}\,\,\, M^D_\nu=U^{\nu \dagger}_L M_\nu U^\nu_R=\operatorname{diag}(m_1\,,\,m_2\,,\,m_3).
\end{equation}
The fields $n_L$ and $n_R$ correspond to the mass basis. Since the right-handed neutrinos remains unobserved, we may, without any loss of generality, assume that they are in a diagonal basis, which amounts to setting $U^\nu_R=I$. As a result, we write $U^\nu_L=U_\text{PMNS}$. Neglecting CP-violating phases, the PMNS matrix is parametrized as follows:
\begin{eqnarray}\small
U_\text{PMNS}=\left(
\begin{array}{ccc}
 \cos \theta  \cos \phi  & \sin \theta  \cos \phi  & \sin \phi  \\
 -\cos \beta  \sin \theta -\sin \beta  \cos \theta  \sin \phi  & \cos \beta  \cos \theta -\sin \beta  \sin \theta  \sin \phi  & \sin \beta  \cos \phi \\
 \cos \beta  \sin \theta -\cos \beta  \cos \theta  \sin \phi  & -\sin \beta \cos \theta -\cos \beta  \sin \theta  \sin \phi  & \cos \beta  \cos \phi  \\
\end{array}
\right),
\end{eqnarray}
which allows us to express the entries of $G^{\nu}$ as
\begin{eqnarray}
    && G^\nu_{11}=\frac{m_1}{v_\Phi} \cos \theta  \cos \phi ,\,\,\,\,\,\,
    G^\nu_{12}=\frac{m_2}{v_\Phi} \sin \theta  \cos \phi ,\,\,\,\,\,G^\nu_{13}=\frac{m_3}{v_\Phi} \sin \phi ,\nonumber \\
    && G^\nu_{22}=\frac{m_2}{v_\Phi} (\cos \beta  \cos \theta -\sin \beta  \sin \theta  \sin \phi ),\nonumber \\
    && G^\nu_{21}=\frac{m_1}{v_\Phi} (-\sin \beta  \cos \theta  \sin \phi -\cos \beta  \sin \theta ),\,\,\,\,\, G^\nu_{23}=\frac{m_3}{v_\Phi} \sin \beta  \cos \phi ,\nonumber \\
    &&G^\nu_{33}=\frac{m_1}{v_\Phi} \cos \theta  \cos \phi,  \,\,\, G^\nu_{31}=\frac{m_1}{v_\Phi} (\cos \beta  \sin \theta -\cos \beta  \cos \theta  \sin \phi ),\nonumber \\
    && G^\nu_{32}=\frac{m_2}{v_\Phi} (-\sin \beta  \cos \theta -\cos \beta  \sin \theta  \sin \phi ).
    \label{couplings}
\end{eqnarray}

Neutrino oscillation experiments yield three mixing angles and two independent mass-squared differences. Neglecting CP-violating phases, the current best-fit values for these parameters are \cite{Esteban:2024eli} 
\begin{eqnarray}
   && |\Delta m_{\mbox{sol}}^2| \approx(7.4 - 7.9)\times 10^{-5}\mbox{eV}^2,\,\,\,\,|\Delta m_{\mbox{atm}}^2|\approx2.5\times 10^{-3}\mbox{eV}^2,\,\,\nonumber \\
   &&\quad \quad \quad \theta \approx 35^o,\,\,\,\, \beta \approx 45^o\,\,\,\,\, \mbox{and}\,\,\,\,\, \phi \approx 8.5^o .
    \label{angles}
\end{eqnarray}

Since neutrino oscillation data do not determine the absolute values of the three neutrino masses, certain assumptions must be adopted to proceed. We set $m_1(m_3)= 0$, for the normal ordering (NO) and inverted ordering (IO) scenarios, respectively. This results in $m_2= \sqrt{\Delta m_{\mathrm{solar}}^2}= 0.86 \times 10^{-2}\mbox{ eV}$ and $m_3 = \sqrt{ \Delta m^2_{\mathrm{atm}}}= 5\times 10^{-2} \mbox{ eV}$ for NO. To evaluate the Yukawa couplings in the NO case for $v_\Phi=1$ eV, we simply substitute the values  $m_1=0$, $m_2 = 0.86 \times 10^{-2}\mbox{ eV}$,  $m_3 = 5\times 10^{-2} \mbox{ eV}$ in \autoref{couplings}. On the other hand, for IO, we have  $m_2= \sqrt{\Delta m_{\mathrm{solar}}^2}+\sqrt{\Delta m_{\mathrm{atm}}^2}=5.86\times 10^{-2} \mbox{ eV}$ and $m_1 = \sqrt{\Delta m_{\mathrm{atm}}^2}=5\times 10^{-2}\mbox{ eV}$. The procedure to obtain the respective Yukawa couplings for $v_\Phi=1$ eV follows analogously to the NO case. 

In this way, the NO and IO scenarios correspond to distinct sets of Yukawa couplings for a given value of $v_\Phi$. In what follows, we assume $v_\Phi = 1$ eV and explore the resulting implications for lepton-flavor-violating processes, specifically $\mu \to e \gamma$ and $\mu \to \bar eee$.

\section{Lepton flavor violation}\label{sec:LFV}
\begin{figure*}[ht!]
    \centering
    \begin{subfigure}[t]{0.49\textwidth}
        \centering
        \includegraphics[width=\linewidth]{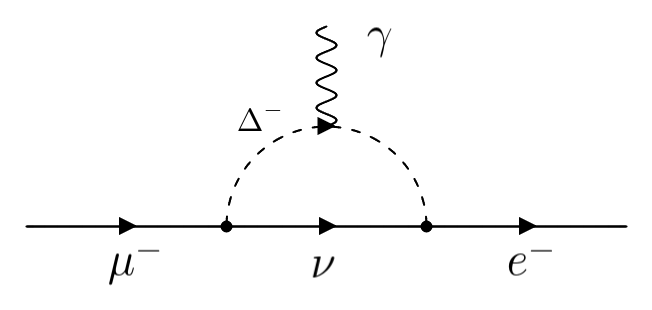}
        \caption{}
        \label{fig:F1}
    \end{subfigure}
    \hfill
    \begin{subfigure}[t]{0.49\textwidth}
        \centering
        \includegraphics[width=\linewidth]{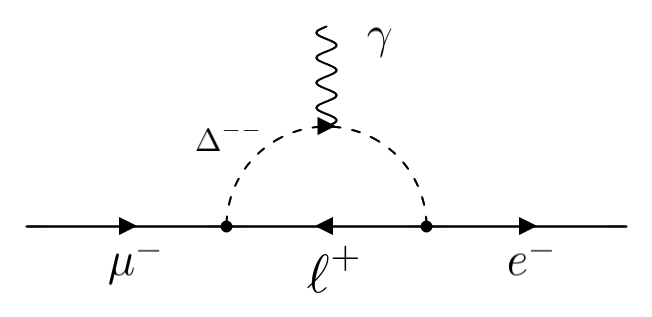}
        \caption{}
        \label{fig:F2}
    \end{subfigure}
    \hfill
    \begin{subfigure}[c]{0.49\textwidth}
        \centering
        \includegraphics[width=\linewidth]{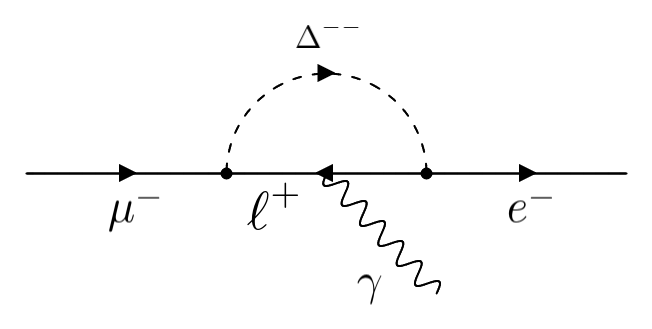}
        \caption{}
        \label{fig:F3}
    \end{subfigure}
    \caption{Feynman diagrams for the process $\mu \to e \gamma$.} 
    \label{fig:mef}
\end{figure*}

\begin{figure}
    \centering
    \includegraphics[width=0.5\linewidth]{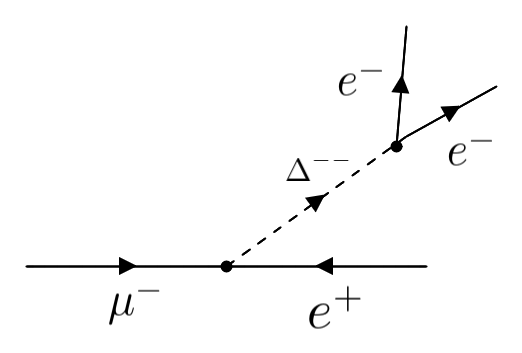}
    \caption{Feynman diagram for the process $\mu \to \bar eee$ .}
    \label{fig:meee}
\end{figure}

As discussed  in \autoref{appendix:A1}, for the values of the parameters $M$ and $\mu_S$ that yield $v_\Phi$ at the eV scale, the scalar spectrum of the sextet decouples from the original scalar content of the 331RHN and acquire masses around $7$ TeV\footnote{Current LHC lower bounds on the mass of doubly charged scalars require $M_{\Delta^{++}}> 1$ TeV \cite{ParticleDataGroup:2024cfk}.}. In this mass regime   the sextet spectrum remains beyond  the current  probing capabilities of the LHC reach at present.  Nevertheless, the charged scalars  within  the sextet may still induce sizable contributions to lepton flavor violating decay processes, such as $\mu \to e \gamma$ (\autoref{fig:mef}) and $\mu \to \bar e ee$ (\autoref{fig:meee}), which we analyze below\footnote{While several lepton flavor violating processes exist, $\mu \rightarrow e \gamma$ and $\mu \to \bar eee$ currently provide the most stringent experimental constraints \cite{Lindner:2016bgg}. For an interesting investigation of this subject in  a particular model, see: \cite{Toma:2013zsa}.}. 

We also emphasize that, since the particle content of the scalar sextet $S$ decouples from the original scalar sector of the 331RHN, and given that the original 331RHN particle content  does not induce rare lepton decays, such processes emerge as a distinctive signature of the extended model, because they are mediated  exclusively by scalars fields contained in the sextet $S$.

The values of the Yukawa couplings, $G^\nu_{ab}$,  related to the NO and IO cases for $v_\Phi=1$ eV are given by
\begin{eqnarray}\small
G^\nu_{NO}\approx \left(\begin{array}{ccc}
 0 & 4.87 \times 10^{-3} & 7.39 \times 10^{-3}  \\
 \\
0 & 4.46 \times 10^{-3} & 3.49 \times 10^{-2} \\
 \\
0 & -5.49 \times 10^{-3} & 0  \end{array}\right),  \
G^\nu_{IO}\approx\left(\begin{array}{ccc}
4 \times 10^{-2} & 3.32 \times 10^{-2}  & 0  \\
 \\
-2.45 \times 10^{-2} &3.04\times 10^{-2} &0 \\
 \\
2.46 \times 10^{-2} & -3.74 \times 10^{-2} & 4 \times 10^{-2} \end{array}\right).
\label{Gmatrix}
\end{eqnarray}

The null entries in the Yukawa matrix arise from the assumption that $m_1 \approx 0$  in the NO scenario and $m_3 \approx 0$ in the IO scenario. As discussed above, within the energy regime relevant for our analysis, the scalar sextet $S$ decouples from the scalar triplets $\eta, \rho$ and $\chi$. Consequently,  the process $\mu \to e\gamma$ is mediated exclusively by the charged scalars $\Delta^+$ and $\Delta ^{++}$, as depicted in the Feynman diagrams shown in \autoref{fig:mef}. Since the spectrum of scalars of $S$ exhibit mass degeneration, the contributions from the charged scalars of the sextet to the  $\mu \to e\gamma$ decay  can be   approximated by the following expression \cite{Lavoura:2003xp}:
\begin{equation}
    Br(\mu \to e\gamma) \approx 27 \alpha  \frac{ |G^\nu_{11}G^\nu_{12} + G^\nu_{12} G^\nu_{22} + G^\nu_{13}G^\nu_{32}|^2 }{ 64 \pi G^2_F M^4_{\Delta^{++}}}.
    \label{BR}
\end{equation}
By substituting the expressions for $M_{\Delta^{++}}$, giving in \autoref{m++}, the parameters $G^\nu_{ij}$ (\autoref{Gmatrix}) and using $\alpha=1/137$, $G_F=1.116 \times 10^{-5}$ GeV$^{-2}$, the mixing angles  in \autoref{angles}, $v_\eta=178$ GeV and $v_{\chi^{\prime}}=10$ TeV, our illustrative case predicts,
\begin{align}
    &{Br}(\mu \to e \gamma) = 3.52 \times 10^{-19} \quad \text{(NO),}\\
    &{Br}(\mu \to e \gamma) = 3.52 \times 10^{-15} \quad \text{(IO),}
\end{align}
for $M=10^{-1}$ keV and $v_\Phi=1$ eV.  The current upper bound on this process is ${Br}(\mu \to e\gamma)< 4.3 \times 10^{-13}$ \cite{MEG:2016leq} with prospect of achieving a sensitivity of  $6\times 10^{-14}$ \cite{Meucci:2022qbh}.   

The Yukawa interactions in \autoref{yukawa2} provide tree level contribution to the muon decay process $\mu \to \bar eee$ as depicted in \autoref{fig:meee}. The current upper bound  for this decay is ${Br}(\mu \to \bar e ee)< 10^{-12}$ \cite{ParticleDataGroup:2024cfk}. The approximated expression to this decay is given by \cite{Lavoura:2003xp}
\begin{equation}
     {Br}(\mu \to \bar e ee) \approx \frac{|G^\nu_{11}|^2 |G^\nu_{12}|^2}{4G_F^2 M^4_{\Delta^{ ++}}} \approx  1.1|G^\nu_{11}|^2 |G^\nu_{12}|\left(\frac{200\mbox{GeV}}{M_{\Delta^{++}}}\right)^4.
\end{equation}
For this rare decay process, our illustrative scenario predicts,
\begin{align}
    &{Br}(\mu \to\bar e ee) = 9.89 \times 10^{-17} \quad \text{(IO)},
\end{align}
while for the NO case the prediction is negligible since $m_1 \approx 0$. Note that our benchmark respects experimental upper bounds, and the IO scenario aligns with the prospective value within  one order of magnitude in the case of the $\mu \to e \gamma$ decay.

\section{Cosmological implications: $\Delta N_\text{eff}$}
\label{sec:Neff}

The effective number of relativistic neutrino species, $N_\text{eff}$, is a key parameter to probe the thermal history of the early Universe. Observations of the Cosmic Microwave Background (CMB) \cite{Planck:2018nkj,Planck:2018vyg} and the predictions of Big Bang Nucleosynthesis \cite{ParticleDataGroup:2024cfk,Cyburt:2015mya} exhibit remarkable agreement with the SM prediction of $N_\text{eff}^\text{SM} = 3.044$ \cite{deSalas:2016ztq,Akita:2020szl,Froustey:2020mcq,Bennett:2020zkv}, corresponding to the three known families of relativistic neutrinos. Any additional contribution to the radiation energy density of the Universe is typically parametrized as
\begin{equation}
\Delta N_\text{eff} \equiv N_\text{eff}^{\text{exp}} - N_\text{eff}^\text{SM},
\end{equation}
where $N_\text{eff}^{\text{exp}}$ denotes the experimentally inferred value. 

The most recent cosmological constraint from the Dark Energy Spectroscopic Instrument (DESI) \cite{DESI:2024mwx} imposes $N_\text{eff}^{\text{exp}} = 3.10 \pm 0.17$ at $1\sigma$ confidence level, leading to an upper bound of $\Delta N_\text{eff} < 0.4$ at $2\sigma$. However, although it is the most recent measurement, the $2\sigma$ constraint from DESI remains weaker than that of the Planck mission, which reports $N_\text{eff}^{\text{exp}} = 2.99^{+0.34}_{-0.33}$ \cite{Planck:2018vyg}. The agreement of these observations with the SM prediction constitutes a cornerstone of the SM's success and serves as a stringent test for its possible extensions.

However, the presence of light particles in SM extensions, such as right-handed neutrinos,  can modify the predicted value of $N_\text{eff}$ \cite{Escudero:2018mvt,Luo:2020sho,Luo:2020fdt,Abazajian:2019oqj,Anchordoqui:2012qu,Calle:2019mxn,Borah:2024twm,Biswas:2021kio}. Furthermore, this quantity can also be affected by non-standard interactions among left-handed neutrinos, as they alter its decoupling temperature. Both effects are significant for constraining SM extensions.

In our model, both new interactions and new light degrees of freedom contribute to an increase in the value of $N_\text{eff}$. The presence of a new neutral gauge boson, $Z^\prime$, introduces extra interactions involving  electrons, positrons and left-handed neutrinos, modifying the decoupling temperature of the $\nu_L$ and impacting $N_\text{eff}$. Besides, as we will show, the right-handed neutrinos in our model reach thermal equilibrium in the early Universe, and depending on their decoupling temperature, they can significantly contribute to $\Delta N_\text{eff}$. 

The relevant interactions are described by the following Lagrangian terms \cite{Pisano:1992bxx,Foot:1994ym}:
\begin{eqnarray}
   && {\cal L}_{W^\prime} \supset\frac{g}{\sqrt{2}}\bar {\nu}^C_{_R} \gamma^\mu e_{_L} W^{\prime}_\mu  + \mbox{H.c.,} \nonumber \\
    && {\cal L}_{U^0} \supset \frac{g}{\sqrt{2}}\bar {\nu}_{_L} U_\text{PMNS}\gamma^\mu \nu^C_{_R} U^{0\dagger}_\mu+ \mbox{H.c.,}\nonumber  \\ 
    &&{\cal L}_{Z^\prime} \supset -\frac{g}{2c_W}\frac{1-2s^2_W}{\sqrt{3-4s^2_W}}\bar \nu_L \gamma^\mu \nu_L Z^{\prime}_\mu -\frac{g}{c_W}\frac{c^2_W}{\sqrt{3-4s^2_W}}\bar \nu_R \gamma^\mu \nu_R Z^{\prime}_\mu\,,
    \label{C-N-currents}
\end{eqnarray}
Additionally, the interactions of the charged leptons ($l=e\,,\,\mu\,,\,\tau$) with the $Z^{\prime}$ boson are given by
\begin{eqnarray}
    {\cal L}_{Z^\prime} \supset - \frac{g}{4c_W \sqrt{3-4 s^2_W}} \bar l \gamma^\mu \left( (3-4c^2_W) + \gamma_5  \right)l Z^{\prime}_{\mu}.
\end{eqnarray}
Finally, interactions with quarks are given by
\begin{eqnarray} \label{eq:4}
\mathcal{L}_{Z'} \supset  &&  - \frac{g}{2c_W} \frac{\sqrt{ 3 - 4 c^2_W} }{3} \left[\overline{u}_{L}  \gamma^{\mu} u_{L} \right]Z^{\prime}_\mu -  \frac{g}{ 2 c_W}\frac{2(1- s_W^2)}{\sqrt{3- 4s_W^2}} \left[\overline{t}_{L}  \gamma^{\mu} t_{L}  \right] Z^{\prime}_{\mu} \nonumber \\ 
 &&   - \frac{g}{2c_W} \frac{\sqrt{3 - 4 s^2_W}}{3} \left[\overline{d}_{L}  \gamma^{\mu} d_{L} \right]Z^{\prime}_\mu -  \frac{g}{ 2 c_W}\frac{2(1- s_W^2)}{\sqrt{3- 4s_W^2}} \left[\overline{b}_{L}  \gamma^{\mu} b_{L}  \right] Z^{\prime}_{\mu} ,
\end{eqnarray}
where $u=(u\,,\,c)^T$ and $d=(d\,,\,s)^T$. 

In \autoref{sec:left_handed}, we calculate the impact of $Z^\prime$ mediated left-handed neutrino annihilation on $N_\text{eff}$. Then, in \autoref{sec:right_handed}, we incorporate the contribution of the right-handed neutrinos to this observable, after analyzing their thermalization. Although the decoupling of right-handed neutrinos is mediated by all non-standard gauge bosons, the relevant amplitudes involving $W'$ and $U^0$ are t-channel processes. They are therefore subdominant compared to $Z'$ mediated s-channel amplitudes, presented in \autoref{app:sterile_neutrino}. Thus, we restrict our analysis to $Z^\prime$. Since the gauge boson masses are related as described in \autoref{sec:general_aspects}, the lower bounds derived for $m_{Z'}$ can be extended to all exotic gauge bosons.

\subsection{Left-handed neutrinos and $Z^\prime$}
\label{sec:left_handed}
In the SM, the left-handed neutrinos decouple from the primordial plasma when weak interactions becomes inefficient, at a temperature $T_\text{dec}^{\nu_L} \sim 1$ MeV. In our extended model, the presence of a new neutral gauge boson, $Z^\prime$, modify the decoupling dynamics. The new interactions modify annihilation processes of left-handed neutrino into electrons and positrons, potentially shifting the decoupling temperature and enhancing the contribution of left-handed neutrinos to $N_\text{eff}$.

In this subsection, we compute the impact of $Z^\prime$-mediated interactions on the decoupling of SM-like left-handed neutrinos\footnote{The contribution from right-handed neutrinos is discussed in \autoref{sec:right_handed}.}. The contribution of the three active neutrino flavors to $N_\text{eff}$ is given by \cite{Biswas:2021kio,EscuderoAbenza:2020cmq}
 \begin{equation}
     N_\text{eff} = 3 \left(\frac{11}{4} \right)^{4/3} \left(\frac{T_{\nu_L}}{T_\gamma} \right)^4\,,
 \end{equation}
where $T_{\nu_L}$ and $T_\gamma$ denote the temperatures of left-handed neutrinos and photons, respectively, at a time after neutrino decoupling.

In order to obtain the evolution of $T_{\nu_L}$ in the early Universe, we solve the Boltzmann equation for the temperature, which can be derived starting from the Liouville equation,
 \begin{equation}\label{Eq:Liouville}
     \frac{\partial f}{\partial t} - H p \frac{\partial f}{\partial p} = \mathcal{C}[f]\,,
 \end{equation}
 where $f$ is the distribution function, $p$ is the momentum, and $H$ is the Hubble expansion rate,
 \begin{equation}
     H = \sqrt{\frac{8 \pi}{3} \frac{\rho_\text{tot}}{m_\text{Pl}^2}}\,,
 \end{equation}
 where $m_\text{Pl} = 1.22 \, \times \, 10^{19}$ GeV is the Planck mass, $\rho_\text{tot}$ is the total energy density of the Universe, and $\mathcal{C}[f]$ is the collision term which can be written for a generic species $\chi$ as
 \begin{align}
     &\mathcal{C}[f_\chi] \equiv -\frac{\mathcal{S}}{E_\chi} \sum_{X\,, Y} \int \prod_i d\Pi_{X_i} \prod_j d\Pi_{Y_j} \left(2\pi \right)^4  \delta^4 \left(p_\chi + p_X - p_Y \right) \times  \nonumber \\
     & |\mathcal{M}|^2 \left[f_\chi \prod_i f_{X_i} \prod_j \left[1 \pm f_{Y_j} \right] -  f_\chi \left[1 \pm f_{\chi} \right]  \prod_j f_{Y_j} \prod_i  \left[1 \pm f_{X_j} \right]\right]\,,
 \end{align}
where $f_\chi\,,f_{X_i}\,, f_{Y_j}$ are the distribution functions of the particles $\chi$, $X_i$, and $Y_j$ respectively. The $+$ sign applies to bosons and the $-$ sign to fermions. $\mathcal{S}$ is the symmetrization factor of $1/2!$ for each pair of identical particles in initial or final states. $d\Pi_X \equiv \frac{g_X}{(2\pi)^3} \frac{d^3p_X}{2E_X}$ represents the phase space of a particle $X$ with $g_X$ internal degrees of freedom. In our case, $\mathcal{S} = 1$, and $g_{\nu_L} = g_{\nu_R} = 2$. The delta function enforces energy-momentum conservation, and $|\mathcal{M}|^2$ is the squared amplitude summed over final spin states and averaged over initial spins of the $\chi + X \leftrightarrow Y$ process\footnote{Assuming T (or CP) invariance, $ |\mathcal{M}|^2_{\chi + X \to Y} = |\mathcal{M}|^2_{Y \to \chi + X }$.}. The collision term includes all processes involving $\chi$  and can be decomposed as 
 \begin{equation}
     \mathcal{C} = \mathcal{C}_\text{ann} +\mathcal{C}_\text{coann} + \mathcal{C}_\text{scat}  + \mathcal{C}_\text{dec}\,,
 \end{equation}
where we explicitly isolate the annihilation, coannihilation, scattering, and decay terms. Other processes, such as three-body final states, are in principle possible. As a reasonable approximation, we consider only the annihilation term \cite{Kawasaki:2000en}. 

Since the energy density of a species $\chi$ is defined as $\rho_\chi \equiv \frac{g_\chi}{(2\pi)^3}\int d^3p \, E_\chi f_\chi$, multiplying the \autoref{Eq:Liouville} by $g_\chi E_\chi p /(2\pi)^3$ and integrating yields \cite{Hannestad:1995rs}
 \begin{equation}
     \frac{d\rho_\chi}{dt} + 3 H \left(\rho_\chi + P_\chi \right) = \frac{\delta \rho_\chi}{\delta t} = g_\chi\int   \frac{d^3 p}{(2 \pi)^3} E_\chi \mathcal{C}[f_\chi]\,,
 \end{equation}
where $P_\chi \equiv \frac{g_\chi}{(2\pi)^3} \int \frac{|p|^2}{3E_\chi} f_\chi d^3p$ is the pressure and $\frac{\delta \rho_\chi}{\delta t}$ represents the energy density transfer rate. For a relativistic particle, we have $P_\chi = \frac{1}{3} \rho_\chi$. As a suitable approximation we neglect chemical potentials and assume $T_\nu = T_{\nu_e} = T_{\nu_\mu} = T_{\nu_\tau}$. As we expect that the left-handed neutrinos just decouple around $T \sim \mathcal{O}(1)$ MeV, we consider temperatures in the range $1 \text{ eV} < T_\gamma < 20 \text{ MeV}$, where $T_\gamma$ is the photon temperature. A thermal bath consisting of electrons, positrons, photons, and neutrinos has total energy density equal to
\begin{equation}
    \rho_\text{tot} = \rho_\gamma + \rho_e + 3 \rho_{\nu_L} \,.
\end{equation}

The energy density transfer rate for the particle with label $1$ in the $1 \, + 2 \, \leftrightarrow \, 3 \, + \, 4$ process, where $T_1 = T_2 = T$ and $T_3 = T_4 = T^\prime$, is \cite{EscuderoAbenza:2020cmq}
\begin{equation}
    \frac{\delta \rho_1}{\delta t}  = \frac{g_1 g_2}{16 \pi^4} \int_{s_\text{min}}^\infty ds \, p_{12}^2 \left(s + m_2^2 - m_1^2 \right) \sigma (s)\left[ T^\prime K_2 \left(\frac{\sqrt{s}}{T^\prime} \right) - T K_2 \left(\frac{\sqrt{s}}{T} \right) \right]\,,
\end{equation}
where $K_2(x)$ is the modified Bessel function of the second kind,   $s_\text{min} = \text{min}\left[(m_1 + m_2)^2 , (m_3 + m_4)^2 \right]$, $p_{12} = \left[s - (m_1 + m_2)^2 \right]^{1/2}\left[s-(m_1-m_2)^2\right]^{1/2}/(2 \sqrt{s})$, and $\sigma(s)$ is the usual cross section summed over final spin states and averaged over initial spins. To proceed further, we simplify the energy transfer rate considering the massless limit,
\begin{equation}
    \frac{\delta \rho_1}{\delta t}  = \frac{g_1 g_2}{16 \pi^4} \int_{0}^\infty ds \, s^2 \, \sigma (s)\left[ T^\prime K_2 \left(\frac{\sqrt{s}}{T^\prime} \right) - T K_2 \left(\frac{\sqrt{s}}{T} \right) \right]\,.
\end{equation}

Using the chain rule, we obtain the evolution equation for the neutrino temperature
\begin{equation}
     \frac{dT_\nu}{dt} = \frac{-12 H \rho_\nu +\frac{\delta \rho_{\nu_e}}{\delta t} + 2 \frac{\delta \rho_{\nu_\mu}}{\delta t}}{3 \frac{\partial \rho_\nu}{\partial T_\nu}}\,.
\end{equation}
Since neither muons nor taus are part of the thermal bath, we conclude that $\frac{\delta \rho_{\nu_e}}{\delta t} \neq  \frac{\delta \rho_{\nu_\mu}}{\delta t}$. However, we assume  $\frac{\delta \rho_{\nu_\mu}}{\delta t} = \frac{\delta \rho_{\nu_\tau}}{\delta t}$, hence the factor of $2$ multiplying $\frac{\delta \rho_{\nu_\mu}}{\delta t}$ in the equation above.

Finally, using the chain rule again, we obtain the temperature evolution for the plasma,
 \begin{equation}
     \frac{dT_\gamma}{dt} = - \frac{4 H \rho_\gamma + 3 H (\rho_e + P_e) + \frac{\delta \rho_{\nu_e}}{\delta t} + 2 \frac{\delta \rho_{\nu_\mu}}{\delta t}}{\frac{\partial \rho_\gamma}{\partial T_\gamma} + \frac{\partial \rho_e}{\partial T_\gamma}}\,.
 \end{equation}
Additionally, following Ref. \cite{EscuderoAbenza:2020cmq}, we  also account for the leading order QED finite temperature corrections through the term $P_\text{int}$,
\begin{equation}\label{eq:Foton}
    \frac{dT_\gamma}{dt} = - \frac{4 H \rho_\gamma + 3 H (\rho_e + P_e) + 3 H T_\gamma \frac{d P_\text{int}}{d T_\gamma} + \frac{\delta \rho_{\nu_e}}{\delta t} + 2\frac{\delta \rho_{\nu_\mu}}{\delta t}}{\frac{\partial \rho_\gamma}{\partial T_\gamma} + \frac{\partial \rho_e}{\partial T_\gamma} + T_\gamma \frac{d^2 P_\text{int}}{d T_\gamma^2}}\,.
\end{equation}
We are not concerned with obtaining the SM contribution to $\frac{\delta \rho_\nu}{\delta t}$ here, as it has already been extensively discussed in previous work. Instead, we directly use the results from \cite{EscuderoAbenza:2020cmq} and compute the non-standard contribution from $Z^\prime$. To summarize, we solve the following system of differential equations:
\begin{align}\nonumber
    \frac{dT_\gamma}{dt} &= - \frac{4 H \rho_\gamma + 3 H (\rho_e + P_e) + 3 H T_\gamma \frac{d P_\text{int}}{d T_\gamma} + \frac{\delta \rho_{\nu_e}}{\delta t} + 2\frac{\delta \rho_{\nu_\mu}}{\delta t} + 3 \frac{\delta \rho_\nu}{\delta t}\Big|_{Z^\prime}}{\frac{\partial \rho_\gamma}{\partial T_\gamma} + \frac{\partial \rho_e}{\partial T_\gamma} + T_\gamma \frac{d^2 P_\text{int}}{d T_\gamma^2}}, \\ 
    \label{eq:active_neutrinos_coupled}
    \frac{dT_{\nu}}{dt} &=- \frac{12 H \rho_\nu - \frac{\delta \rho_{\nu_e}}{\delta t} - 2 \frac{\delta \rho_{\nu_\mu}}{\delta t} - 3\frac{\delta \rho_\nu}{\delta t}\Big|_{Z^\prime}}{3 \frac{\partial \rho_\nu}{\partial T_\nu}}\, ,
\end{align}
where we isolate $\frac{\delta \rho_\nu}{\delta t}\Big|_{Z^\prime}$, denoting the energy density transfer rate due to $Z^\prime$. The relevant process then is $\nu_L + \bar \nu_L \leftrightarrow e^- +  e^+$, so we write
\begin{equation} \label{eq:ZprimeDeltarho}
    \frac{\delta \rho_\nu}{\delta t}\Big|_{Z^\prime} = \frac{g_1 g_2}{64 \pi^4} \int_{0}^\infty ds \, s^2 \, \sigma_{\nu \bar \nu \to e^- e^+}^{Z^\prime} (s)\left[ T_\gamma K_2 \left(\frac{\sqrt{s}}{T_\gamma} \right) - T_\nu K_2 \left(\frac{\sqrt{s}}{T_\nu} \right) \right],
\end{equation}
where $\sigma_{\nu \bar \nu \to e^- e^+}^{Z^\prime}$ is the $Z^\prime$ induced cross section. 
As detailed in \autoref{appendix:A}, this cross section is
 \begin{align}\nonumber
    \sigma_{\nu \bar \nu \to e^- \bar e^+}^{Z^\prime} & =  \frac{g^4 s  \cos^2 2 \theta_W  (\cos 4 \theta_W  -2 \cos 2 \theta_W+2) }{768 \pi \cos^4 \theta_W  (2 \cos 2 \theta_W +1)^2 m_{Z^\prime}^4} \\ 
    & +\frac{g^4 s \cos 2 \theta_W (\cos 4 \theta_W  -2 \cos 2 \theta_W+2) }{384 \pi  \cos^4 \theta_W (2 \cos 2 \theta_W +1) m_{Z}^2 m_{Z^\prime}^2} \nonumber \\ 
    & +\frac{g^4 \cos ^22\theta_W  \left(m_W^2 + 2 s\right)}{128 \pi \cos^2 \theta_W (2 \cos 2\theta_W+1) m_{Z^\prime}^2 m_W^2}\,.
\end{align}

Due to the exponential fall-off of the integrand in \autoref{eq:ZprimeDeltarho}, and as $m_{Z^\prime} \gg m_Z \,, m_W$, we conclude that the dominant contribution to $\frac{\delta \rho_\nu}{\delta t}\Big|_{Z^\prime}$ comes from the interference terms between the standard gauge boson $Z$ and the new $Z^\prime$.

\begin{figure}
	\centering
	\includegraphics[width=0.5\textwidth]{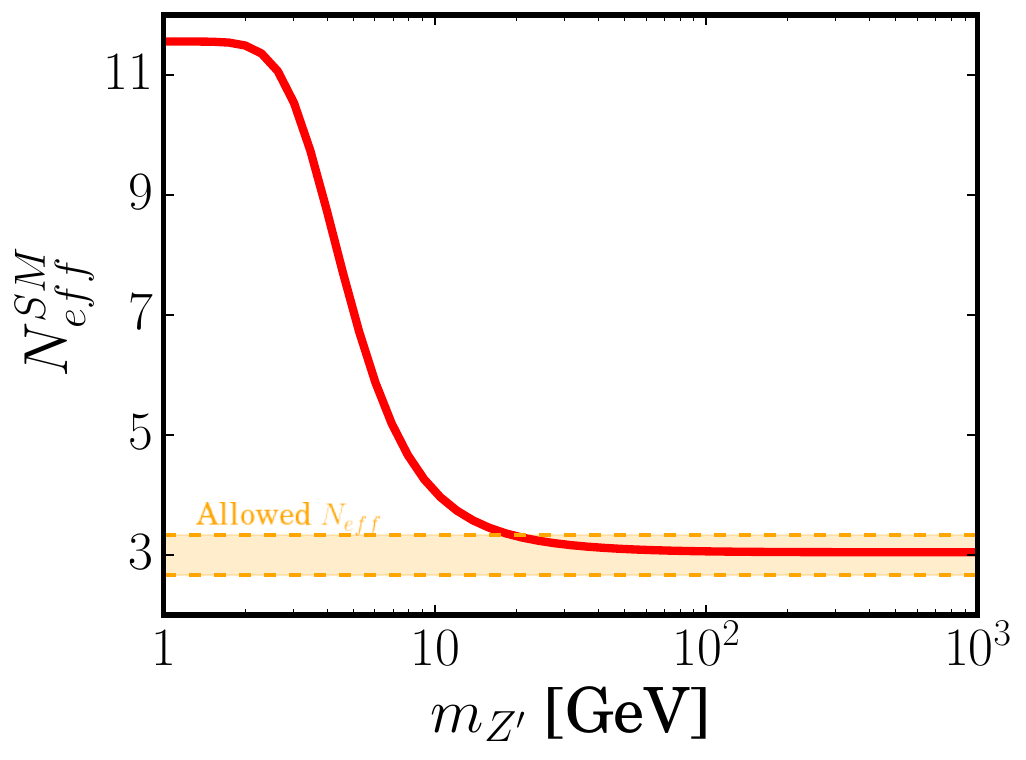}
	\caption{Left-handed neutrino contribution to the effective number of relativistic species as a function of $m_{Z^\prime}$. The orange region represents the allowed limit $N_\text{eff}^{\text{exp}} = 2.99^{+0.34}_{-0.33}$ imposed by Planck. This aspect of the thermal history requires $m_{Z^\prime}\geq 21$ GeV.}
	\label{fig:Neff_activenu}
\end{figure}

We solve the Boltzmann equations for the temperature by modifying the \texttt{NUDEC\_BSM} code \cite{Escudero:2018mvt}, which incorporates the SM case. After implementing the $Z^\prime$ contribution, we obtain the results shown in \autoref{fig:Neff_activenu}.
The red curve shows the evolution of SM-like left-handed neutrino contribution to $N_\text{eff}$ as a function of $m_{Z^\prime}$. The orange band represents the region allowed by Planck: $N_\text{eff}^{\text{exp}} = 2.99^{+0.34}_{-0.33}$. We observe that for $m_{Z^\prime} > 21$ GeV, the $Z^\prime$ contribution remains consistent with this bound\footnote{The contribution from right-handed neutrinos imposes more restrictive constraints; see \autoref{sec:right_handed}.}.
$N_\text{eff}$ remains approximately constant for $m_{Z^\prime} \sim \mathcal{O}(1)$ GeV, which is the regime in which the $Z^\prime$ interaction is strong enough to maintain thermal equilibrium when $T_\gamma = 1$ eV. This constant value, $N_\text{eff} \simeq 11.5$,  can be extracted from $N_\text{eff} = 3 \left( \frac{11}{4}\right)^{4/3} \left( \frac{T_{\nu_L}}{T_\gamma}\right)^4$, with $T_{\nu_L}= T_\gamma$.
As the value of $m_{Z^\prime}$ increases, its contribution becomes increasingly suppressed, and  $N_\text{eff}$ gradually converges to the SM prediction of $3.044$.

\subsection{Right-handed neutrinos}
\label{sec:right_handed}
We previously showed  the effect of $Z'$ on the decoupling of SM-like left-handed neutrinos. Of course, a phenomenologically sound analysis must include the effect of all relativistic species in the plasma. In what follows, we consider the thermal history considering also the presence of the right-handed neutrino population.

First, we need to determine the range of masses for $Z^\prime$ that allows the right-handed neutrinos to reach thermal equilibrium. The right-handed neutrinos maintain thermal equilibrium at some moment of the early Universe as long as $\Gamma(T) > H(T)$. Here, $\Gamma(T)$ represents the interaction rate, given by  $\Gamma(T)\equiv n_{\nu_R} \langle \sigma v \rangle$, where $\langle \sigma v \rangle$ is the thermally averaged cross section:
\begin{equation}
    \langle \sigma v \rangle = \frac{T}{32 \pi^4 n_{\nu_R}^2} \int_{0}^\infty d\mathrm{s} \, \sigma(\mathrm{s}) \, \mathrm{s}^{3/2} K_1 \left( \frac{\sqrt{\mathrm{s}}}{T} \right)\,.
\end{equation}
In this expression $K_1(x)$ denotes the modified Bessel function of the first kind, $\sigma(\mathrm{s})$ is the cross section, and $\mathrm{s}$ is the Mandelstam variable. The factor $n_{\nu_R}$ corresponds to the number density of right-handed neutrinos,
\begin{equation}
    n_{\nu_R} = \frac{3}{4} \frac{\zeta(3)}{\pi^2}g_{\nu_R} T^3\,,
\end{equation}
where $\zeta$ is the Riemann zeta function satisfying $\zeta(3) = 1.20206$.
\begin{figure}
    \centering
    \includegraphics[width=0.7\linewidth]{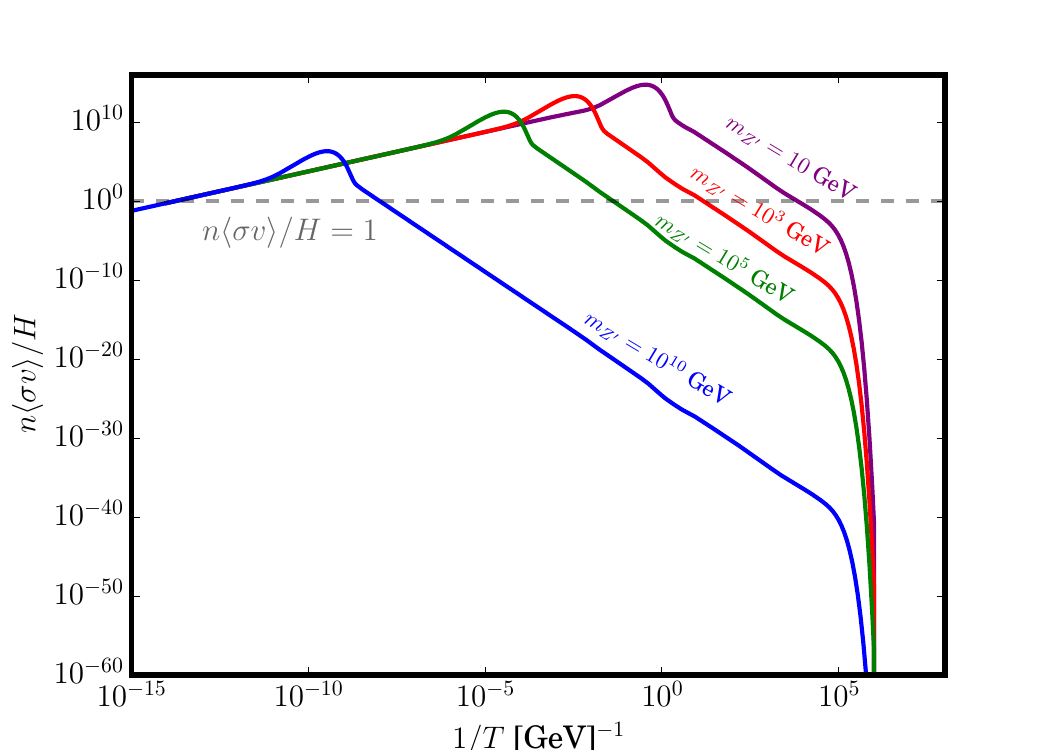}
    \caption{Ratio of the rate of annihilation of right-handed neutrinos and the expansion rate as a function of the inverse temperature for different masses of $Z'$. The dashed gray line represents the unit value, characteristic of decoupling from the SM plasma.}
    \label{RateMass}
\end{figure}

We computed the interaction rate numerically using \texttt{Cuba} \cite{Hahn:2004fe}, neglecting the $t$-channel diagrams mediated by $W^\prime$ and $U^0$, as their contributions are negligible\footnote{For more details, see \autoref{ap.:D}.}. We are left with 3 possible final states: $\nu_R \bar{\nu}_R \to \nu_L \bar{\nu}_L$, $\nu_R \bar{\nu}_R \to l \bar{l}$, and $\nu_R \bar{\nu}_R \to q \bar{q}$, all mediated by $Z^\prime$ in $s$-channels. The squared amplitudes and cross sections for these processes are detailed in \autoref{app:sterile_neutrino}.  The result of $\Gamma / H$ as a function of the inverse temperature is shown in \autoref{RateMass} for different values of $m_{Z^\prime}$. It is evident that right-handed neutrinos are in thermal equilibrium whenever $m_{Z^\prime} < 10^{16}$ GeV.

The thermally averaged cross sections add up to an effective one defined as
\begin{equation}
    \langle \sigma v \rangle = \sum_{X=\nu_L, l, q_1, q_2, q_3} \langle \sigma_{\nu_R \bar{\nu}_R \to X \bar{X}} v \rangle\,,
\end{equation}
where $q_{1,2,3}$ represents the quarks from the first, second, and third families, respectively. Then, the decoupling temperature for right-handed neutrinos, defined implicitly by $\Gamma(T_{\nu_R}^\text{dec}) = H(T_{\nu_R}^\text{dec})$, is
\begin{equation}\label{eq:T_dec_sterile}
    T_{\nu_R}^\text{dec} =  \left(\frac{53}{5}\right)^{1/6} \left(\frac{6 \pi ^4 m_{Z^\prime}^4 \zeta (3) (2 \cos (2 \theta_W)+1)^2}{g^4 m_\text{Pl} (68 \cos (2 \theta_W)+61 \cos (4 \theta_W)+123)} \right)^{1/3}\,,
\end{equation}
where we assume $g_\rho(T_{\nu_R}) = 106.75$, where $g_\rho(T)$ is the effective number of degrees of freedom in energy \cite{Husdal:2016haj}. The expression above is consistent with our numerical results.

Next, we determine $\Delta N_\text{eff}$. Consider the radiation energy density of the Universe in the presence of three right-handed neutrinos after $e^\pm$ annihilation. The total radiation energy density can be expressed as \cite{Anchordoqui:2012qu,Calle:2019mxn}
\begin{align}
    \rho_r &= \rho_\gamma + N_{\nu_L} \rho_{\nu_L} + N_{\nu_R} \rho_{\nu_R} \\ 
           &= \left( 1 + \left(\frac{T_{\nu_L}}{T_\gamma}\right)^4 \left[ N_{\nu_L} + N_{\nu_R} \left(\frac{T_{\nu_R}}{T_{\nu_L}}\right)^4 \right] \right) \rho_\gamma \\ 
           &= \left( 1 + \frac{7}{8}\left(\frac{4}{11}\right)^{4/3} N_\text{eff} \right) \rho_\gamma,
\end{align}
where $\rho_\gamma$ is the energy density of photons, $\rho_{\nu_L}$ and $\rho_{\nu_R}$ are the energy densities of left- and right-handed neutrinos, $T_\gamma$, $T_{\nu_L}$, and $T_{\nu_R}$ are the species temperatures, and $N_{\nu_L} = N_{\nu_R} = 3$ are the number of neutrino flavors.

Here, $N_\text{eff}$ encapsulates the effective number of relativistic species contributing to the radiation energy density. It is defined as
\begin{equation}
    N_\text{eff} = N_{\nu_L} + N_{\nu_R} \left( \frac{T_{\nu_R}}{T_{\nu_L}} \right)^4,
\end{equation}
so the deviation $\Delta N_\text{eff}$ caused by the presence of right-handed neutrinos is found to be
\begin{equation}
    \Delta N_\text{eff} = N_{\nu_R} \frac{T_{\nu_R\,,0}^4}{T_{\nu_L\,,0}^4},
\end{equation}
where $T_{\nu_R\,,0}$ and $T_{\nu_L\,,0}$ are the present-day temperatures of the right- and left-handed neutrinos, respectively.

To proceed further, we relate the neutrino temperatures $T_{\nu_R}$ through entropy conservation. The right-handed neutrinos decouple earlier than the left-handed neutrinos, when the number of relativistic degrees of freedom is $g_s(T_{\nu_R}^\text{dec})$. After its decoupling, the right-handed neutrinos do not experience the entropy release from $e^\pm$ annihilation, leading to a posterior cooling compared to left-handed neutrinos, given by
\begin{equation}
    \frac{T_{\nu_R}}{T_{\nu_L}} = \left(\frac{g_s(T_{\nu_L})}{g_s(T_{\nu_R}^\text{dec})}\right)^{1/3},
\end{equation}
where $g_s(T_{\nu_L}) = 10.75$ after $e^\pm$ annihilation. Substituting this relation into the expression for $\Delta N_\text{eff}$, we obtain,
\begin{equation}
    \Delta N_\text{eff} = N_{\nu_R} \left( \frac{T_{\nu_R}}{T_{\nu_L}} \right)^4 = N_{\nu_R} \left( \frac{10.75}{g_s(T_{\nu_R}^\text{dec})} \right)^{4/3}.
\end{equation}
Inserting $N_{\nu_R} = 3$, the expression simplifies to \cite{Borah:2024twm},
\begin{equation}
    \Delta N_\text{eff} = 0.047 \times 3 \left( \frac{106.75}{g_s(T_{\nu_R}^\text{dec})} \right)^{4/3},
\end{equation}
where $g_s(T_{\nu_R}^\text{dec})$ corresponds to the number of entropic degrees of freedom at the decoupling temperature of the right-handed neutrinos \cite{Husdal:2016haj}.

The Planck 2018 results impose a stringent upper bound on $\Delta N_\text{eff}$, constraining it to $\Delta N_\text{eff} < 0.285$ at the $2\sigma$ confidence level \cite{Planck:2018nkj,Planck:2018vyg}. More recently, the DESI has reported constraints of $\Delta N_\text{eff} < 0.225$ at $1\sigma$ and $\Delta N_\text{eff} < 0.4$ at $2\sigma$. These bounds currently represent the most up-to-date observational limits on $N_\text{eff}$.

\begin{figure}
	\centering
	\includegraphics[width=0.7\textwidth]{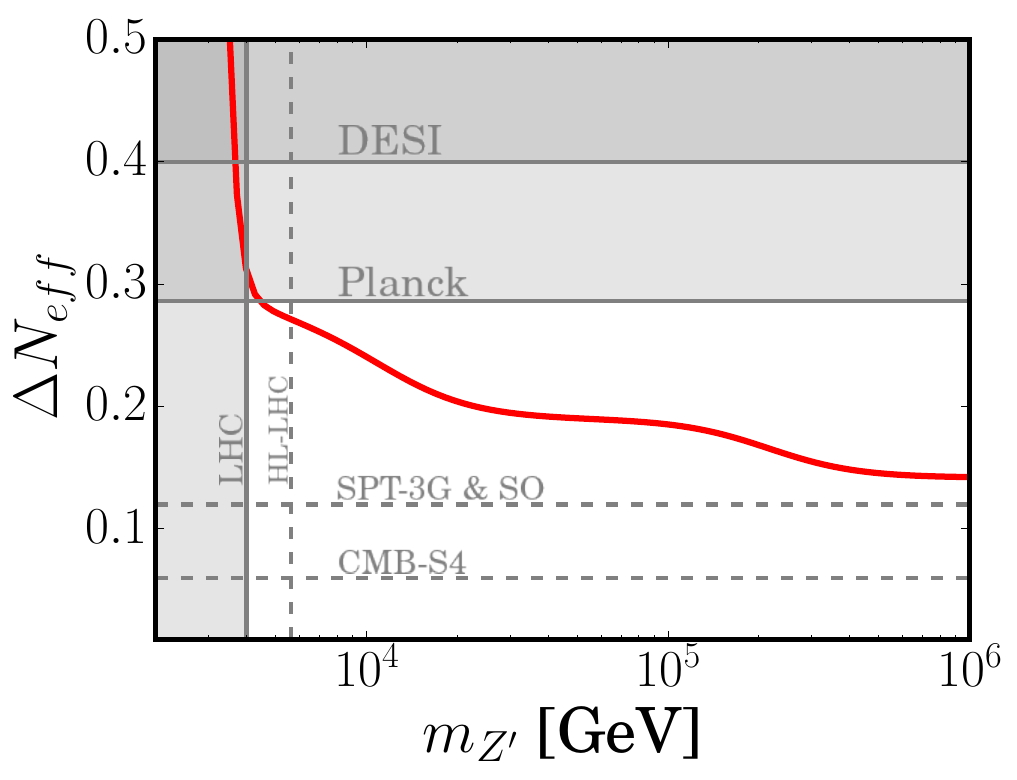}
	\caption{$N_\text{eff}$ evolution in function of $m_{Z^\prime}$.  We conclude that $m_{Z^\prime}> 4.4$ TeV in order to satisfy current limits. Lower bounds from colliders are also displayed, corresponding to $m_{Z^\prime} \geq 4$ TeV (LHC) and  $m_{Z^\prime} \geq 5.6$ TeV (HL-LHC).}
	\label{fig:Neff_sterilenu}
\end{figure}

Future cosmological observations are expected to significantly enhance the sensitivity to $\Delta N_\text{eff}$. The South Pole Telescope (SPT-3G) and the Simons Observatory (SO) are projected to probe deviations as small as $\Delta N_\text{eff} < 0.12$ at the $2\sigma$ level \cite{SPT-3G:2014dbx,SimonsObservatory:2018koc}, providing a more precise determination of potential contributions from light, thermally decoupled species such as right-handed neutrinos. Furthermore, the next-generation experiment CMB-S4 is expected to achieve an even more remarkable sensitivity, with $\Delta N_\text{eff} < 0.06$ at $2\sigma$ \cite{Abazajian:2019eic,CMB-S4:2016ple}.

By comparing our predictions, shown in \autoref{fig:Neff_sterilenu}, with the observational constraints, we conclude that the mass of the $Z^\prime$ boson must satisfy $m_{Z^\prime} > 4.4$ TeV to remain within the allowed region.  This lower bound ensures that the right-handed neutrino decouples early enough in order to prevent excessive contributions to $\Delta N_\text{eff}$ that would violate experimental limits. This lower bound on $m_{Z^\prime}$ translates into  $v_{\chi^{\prime}}> 9.9$ TeV. Such bound is slightly above the one in Ref.~\cite{Coutinho:2013lta}.

In addition to the lower mass bounds derived from the effective number of neutrino species, we juxtapose collider bounds on the $Z^\prime$ mass derived using LHC data, and projections for HL-LHC \cite{Coutinho:2013lta,Alves:2022hcp}. Concretely, we also plot in \autoref{fig:Neff_sterilenu} the LHC result $m_{Z^\prime} \geq 4$ TeV, and the HL-LHC expected lower bound $m_{Z^\prime} \geq 5.6$ TeV (in case of null observation). Furthermore, additional constraints on the $Z^\prime$ mass can arise, for instance, from LEP-II limits \cite{ALEPH:2013dgf}. However, these constraints are model-dependent, and a detailed analysis lies beyond the scope of this work. Nonetheless, according to the literature, such bounds are typically below $\mathcal{O}(1)$ TeV, and therefore weaker than those from $N_\text{eff}$ and LHC.

Our result highlights the sensitivity of $\Delta N_\text{eff}$ to the mass of the $Z^\prime$ boson, offering a valuable probe for testing the viability of this model in light of current and future cosmological observations. However, this constraint could be relaxed in cosmological scenarios where the reheating temperature of the Universe is lower than the decoupling temperature of the right-handed neutrinos \cite{Kawasaki:2000en}. In such cases, the right-handed neutrinos would not thermalize, and their contribution to $\Delta N_\text{eff}$ would be negligible, allowing the $Z^\prime$ mass to fall below the derived limit.

\section{Conclusion}
\label{sec:conclusion}

In this work, we implemented the type-II seesaw mechanism for Dirac neutrino masses within the framework of the  $\text{SU}(3)_L \times \text{U}(1)_N$ model with right-handed neutrinos by introducing a sextet of scalars to the original scalar content of the model. Our model employs a $Z_2$ symmetry to suppress potentially dangerous terms involving quark flavors. After adding a soft term that breaks this $Z_2$ symmetry, a type-II seesaw mechanism for Dirac neutrino masses naturally emerges.  As is common in neutrino mass generation mechanisms, significant implications for lepton flavor violation naturally arise. To assess these effects, we examine our proposed mechanism in the context of the $\mu \to e \gamma$ and $\mu \to \bar{e} ee$ processes. Our analysis indicates that the current experimental upper bounds on these decay channels are consistent with the realization of the type-II seesaw mechanism at a low-energy scale

The presence of right-handed neutrinos and a $Z^{ \prime}$ field induces sizable changes in the effective number of relativistic neutrino species, $N_{\text{eff}}$. New interactions can thermalize the right-handed neutrinos in the early Universe, increasing the radiation energy density and this affecting $N_{\text{eff}}$. We computed the thermalization and decoupling conditions for right-handed neutrinos and derived an explicit dependence of $\Delta N_\text{eff}$ on the mass of the $Z^\prime$ boson. This results into a lower  bound of $m_{Z^\prime} > 4.4$ TeV, which translates into a lower bound on the SSB scale: $v_{\chi^{\prime}} > 9.9$ TeV. This limit is slightly stronger than current LHC constraints based on dilepton searches (See Table III of \cite{Alves:2022hcp}). We remark that this lower bound ensures $Z^\prime$-mediated interactions decouple early enough to prevent right-handed neutrinos from contributing significantly to $\Delta N_\text{eff}$. 
Thus, $N_{\text{eff}}$ can indeed provide a complementary and independent constraint to those obtained at colliders. More broadly, our work highlight that realizations of type-II seesaw mechanisms for Dirac neutrino within extended gauge symmetries offer a compelling interplay between cosmology and neutrino physics, leading to bounds that can exceed those derived from high-energy hadron collider experiments. 

We finalize this conclusion emphasizing that the bound on $m_{Z^\prime}$ inferred from $N_\text{eff}$ is independent of the nature of neutrinos.

\section*{Acknowledgments}
Some of us thank UFPB members for their hospitality, where the initial steps of this work were taken. L. A. was supported by CAPES under grant 88887.827404/2023-00. P. E. is directly funded by CNPq 151612/2024-2. V. O. is  directly funded by FCT through the doctoral program grant with the reference PRT/BD/154629/2022 (\url{https://doi.org/10.54499/PRT/BD/154629/2022}).  V. O. ~also acknowledges support  by the COST Action CA21106 (Cosmic WISPers).
C. A. S. P.  was supported by the CNPq research grants No. 311936/2021-0. 
F. S. Q. is supported by Simons Foundation (Award Number:1023171-RC), FAPESP Grant 2018/25225-9, 2021/01089-1, 2023/01197-4, ICTP-SAIFR FAPESP Grants 2021/14335-0, CNPq Grants 307130/2021-5, and ANID-Millennium Science Initiative Program ICN2019\textunderscore044, and FINEP under the project 213/2024.

\appendix
\section{Higgs spectrum}
\label{appendix:A1}
We briefly develop the scalar sector of the 331RHN considering the hierarchy  $v_\Phi \ll f \ll v_\text{ew} \ll v_{\chi^{\prime}}$, where $v_\eta = v_\rho=\frac{v_\text{ew}}{\sqrt{2}}$, with $v_\text{ew}=246$ GeV being the standard VEV. In this case, the mass matrix for the  CP-even scalars in the  basis $(R_\eta, R_\rho, R_{\chi^{\prime}},  R_\Phi)$ is\\
\begin{frame}

\resizebox{\linewidth}{!}{%
$\displaystyle
M^2_R=\left(
\begin{array}{cccc}
2\lambda_{2}v_{\eta}^2  +\frac{f v_{\rho} v_{\chi}}{2 v_{\eta}}+ \frac{\sqrt{2}M v_{\Phi} v_{\chi}}{4 v_{\eta}} & 
\lambda_{6}v_{\eta} v_{\rho} -\frac{f v_{\chi}}{2} & 
\lambda_{4}v_{\eta} v_{\chi} -\frac{f v_{\rho}}{2} - \frac{\sqrt{2} M v_{\Phi}}{4} & 
\lambda_{12}v_{\eta}v_{\Phi} +\frac{\lambda_{15}v_{\eta}v_{\Phi}}{2} - \sqrt{2} M v_{\chi} \\
\lambda_{6}v_{\eta} v_{\rho} -\frac{f v_{\chi}}{2} &
2 \lambda_{3}v_{\rho}^{2} +\frac{f v_{\eta} v_{\chi}}{2 v_{\rho}} &
\lambda_{5}v_{\rho} v_{\chi}  -\frac{f v_{\eta}}{2}& 
\lambda_{13} v_{\rho} v_{\Phi} \\
\lambda_{4}v_{\eta} v_{\chi} - \frac{f v_{\rho}}{2}-\frac{\sqrt{2} M v_{\Phi}}{4} & 
\lambda_{5}v_{\rho} v_{\chi} -\frac{f v_{\eta}}{2} & 
2\lambda_1 v_{\chi}^{2}  + \frac{ f v_{\eta} v_{\rho}}{2v_{\chi} }+\frac{\sqrt{2} M v_{\eta} v_{\Phi} }{4 v_\chi} &
\lambda_{14}v_{\Phi} v_{\chi}+ \frac{\lambda_{17}v_{\Phi} v_{\chi}}{2}-\frac{\sqrt{2} M v_{\eta}}{4} \\
\lambda_{12}v_{\eta} v_{\Phi} +\frac{\lambda_{15}v_{\eta} v_{\Phi}}{2}- \frac{ \sqrt{2} M v_{\chi}}{4}& 
\lambda_{13} v_{\rho} v_{\Phi}  & 
\lambda_{14}v_{\Phi} v_{\chi} + \frac{\lambda_{17}v_{\Phi} v_{\chi}}{2}-\frac{ \sqrt{2} M v_{\eta}}{4} & 
2\lambda_{10} v_{\Phi}^2 +\lambda_{11} v_{\Phi}^2+\frac{M v_{\eta} v_{\chi}}{2 \sqrt{2} v_{\Phi}}
\end{array}\right).
$}
\end{frame} \\
For our scenario, the component  $R_\Phi$ practically decouples from the other CP-even scalars and has  a mass term equal to $m^2_{H^{\prime}}=\frac{1}{4}\frac{M v_\text{ew}  v_{\chi}}{ v_{\Phi}}$,  where we identify  $R_\Phi \equiv H^{\prime}$. 

The mass matrix of the CP-odd scalars in the basis $(I_{\eta} , I_{\rho} , I_{\chi^{\prime}} , I_{\Phi})$ is
\begin{center}
$M_I^2=\left(\begin{array}{cccc}
\frac{f v_{\rho} v_{\chi}}{2 v_{\eta}}+ \frac{\sqrt{2} M v_{\Phi} v_{\chi} }{4 v_{\eta}} & 
\frac{f v_{\chi}}{2} & 
\frac{f v_{\rho}}{2}+ \frac{\sqrt{2} M v_{\Phi}}{4} & 
-\frac{M v_{\chi}}{2 \sqrt{2}} \\ 
\frac{f v_{\chi}}{2} & 
\frac{f v_{\eta} v_{\chi}}{2 v_{\rho}} & 
\frac{f v_{\eta}}{2} &
0 \\ 
\frac{f v_{\rho}}{2} + \frac{\sqrt{2} M v_{\Phi}}{4} & 
\frac{f v_{\eta}}{2} & 
\frac{f v_{\rho}v_{\eta}}{2 v_{\chi}} +\frac{\sqrt{2} M v_{\Phi} v_{\eta}}{4 v_{\chi}} &
-\frac{M v_{\eta}}{2 \sqrt{2}} \\ 
-\frac{M v_{\chi}}{2 \sqrt{2}} &
0 &
-\frac{M v_{\eta}}{2 \sqrt{2}} & 
\frac{M v_{\eta} v_{\chi}}{2 \sqrt{2} v_{\Phi}}
\end{array}\right)$.
\end{center}
Observe that $I_\Phi $ also decouples from the other CP-odd scalars and acquire mass $m^2_{A^{\prime}}=\frac{1}{4}\frac{M v_\text{ew} v_{\chi}}{ v_{\Phi}}$ where we identify $I_\Phi \equiv A^{\prime}$. Diagonalizing the remaining $3 \times 3$ matrix for $I_\eta , I_{\rho} , I_{\chi^{\prime}} $, we obtain two Goldstone bosons that will be eaten by $Z$ and $Z^{\prime}$. The last pseudo-scalar, $A$, which is a combination of $I_\eta$ and $I_\rho$,  has mass $m^2 _A=\frac{f v_{\chi^{\prime}}}{2}$, see Ref.~\cite{Pinheiro:2022bcs}.

All inert neutral  scalars are bileptons (carrying $L=-2$). Their mass matrix on the basis $(\eta^{0^{\prime}}, \chi^{0},\Delta^{0},\sigma)$ reads

\begin{frame}

\resizebox{\linewidth}{!}{%
$\displaystyle
M_{R^\prime} =\left(
\begin{array}{cccc}
\lambda_{7}v_{\chi}^{2} + \frac{f v_{\rho} v_{\chi}}{v_{\eta}}+ \frac{\sqrt{2}M v_{\Phi} v_{\chi}}{2 v_{\eta}} & 
\lambda_{7}v_{\eta} v_{\chi} + f v_{\rho} - \frac{M v_{\Phi}}{\sqrt{2}} & 
\frac{\lambda_{15}v_{\eta} v_{\Phi} }{\sqrt{2}} & 
 \frac{\lambda_{15}v_{\eta} v_{\Phi} }{\sqrt{2}}-M v_{\chi}  \\ 
\lambda_{7} v_{\eta}v_{\chi} +f v_{\rho} - \frac{M v_{\Phi}}{\sqrt{2}}& 
\lambda_{7}v_{\eta}^{2} +\frac{f v_{\rho} v_{\eta}}{v_{\chi}} + \frac{\sqrt{2} M v_{\Phi}v_{\eta}}{2 v_{\chi}}  & 
\frac{ \lambda_{17}v_{\Phi} v_{\chi}}{\sqrt{2}}-M v_{\eta}  &
 \frac{ \lambda_{17}v_{\Phi} v_{\chi}}{\sqrt{2}}\\ 
 \frac{\lambda_{15}v_{\eta} v_{\Phi}  }{\sqrt{2}} & 
\frac{\lambda_{17}v_{\Phi} v_{\chi} }{\sqrt{2}}-Mv_{\eta} & 
 \lambda_{11}v_{\Phi}^2  + \frac{\lambda_{15}v_{\eta}^2}{2}  - \frac{\lambda_{17}v_{\chi}^2}{2} +\frac{\sqrt{2} M v_{\eta} v_{\chi}}{2v_{\Phi}} & 
\lambda_{11}v_{\Phi}^{2} \\ 
\frac{\lambda_{15}v_{\eta} v_{\Phi}}{\sqrt{2}}-M v_{\chi}&
\frac{\lambda_{17}v_{\Phi} v_{\chi}}{\sqrt{2}}&
 \lambda_{11}v_{\Phi}^{2} & 
\lambda_{11}v_{\Phi}^2  - \frac{\lambda_{15}v_{\eta}^2 }{2} + \frac{ \lambda_{17} v_{\chi}^2}{2} +\frac{\sqrt{2} M v_{\eta} v_{\chi}}{2 v_{\Phi}}
\end{array}\right).$
}
\end{frame} \\ \\
Here we have  two decouplings:  $\Delta ^0$ and  $\sigma^0$  decouple from $\chi^0$ and $\eta^{\prime 0}$, and $\Delta^{0}$ decouples from $\sigma$. In this case, $\Delta^0$ and $\sigma$ are degenerate in mass: $m^2_{\Delta^0} \approx m^2_{\sigma}\approx \frac{1}{4}\frac{M v_\text{ew} v_{\chi}}{ v_{\Phi}} $. 

The singly charged  scalars $ \Delta^+, \rho^{\prime+}$ and $\chi^+$, are bileptons. Their mass matrix in the basis $(\Delta^+, \rho^{\prime+},\chi^+)$ is
\[
M_{h^+}^2=\left(
\begin{array}{ccc}
 \frac{\lambda _{16} v_{\rho }^2 v_{\Phi }-\lambda _{17} v_{\chi }^2 v_{\Phi }+2 M v_{\eta } v_{\chi }}{4 v_{\Phi }} & 
 \frac{1}{4} \lambda _{16} v_{\rho } v_{\Phi } & 
 \frac{1}{4} \left(\lambda _{17} v_{\chi } v_{\Phi }-2 M v_{\eta }\right)  \\
 \frac{1}{4} \lambda _{16} v_{\rho } v_{\Phi } &
 \frac{1}{4} \left(2 \lambda _8 v_{\chi }^2+\lambda _{16} v_{\Phi }^2+\frac{2 \sqrt{2} f v_{\eta } v_{\chi }}{v_{\rho }}\right) & 
 \frac{1}{2} \left(\lambda _8 v_{\rho } v_{\chi
   }+\sqrt{2} f v_{\eta }\right)\\
 \frac{1}{4} \left(\lambda _{17} v_{\chi } v_{\Phi }-2 M v_{\eta }\right) & 
 \frac{1}{2} \left(\lambda _8 v_{\rho } v_{\chi }+\sqrt{2} f v_{\eta }\right) & 
 \frac{
 2\lambda _8 v_{\rho }^{2} v_{\chi }
- \lambda _{17} v_{\chi } v_{\Phi }^2
 + 2\sqrt{2}f v_{\rho }v_{\eta }+2 M v_{\eta } v_{\Phi }}
 {4 v_{\chi }}
\end{array}
\right).
\] \\
Here, $\Delta^+$ also decouples from the other charged scalars and acquires mass $m^2_{\Delta^+} \approx \frac{1}{4}\frac{M v_\text{ew} v_{\chi}}{ v_{\Phi}}$.

The other set of singly charged  scalars  does not carry lepton number, and considering the basis $(\Phi^+, \eta^+,\rho^+)$, we express  their mass matrix as
\[
M_{h^{\prime+}}^2=\left(
\begin{array}{ccc}
 \frac{-\lambda _{15} v_{\eta }^2 v_{\Phi}+\lambda _{16} v_{\rho }^2 v_{\Phi }+2 M v_{\eta } v_{\chi }}{4 v_{\Phi }} &
 \frac{1}{4} \left(\lambda _{15} v_{\eta } v_{\Phi }-2 M v_{\chi }\right) & 
 \frac{1}{4} \lambda _{16} v_{\rho } v_{\Phi } \\
 
  \frac{1}{4} \left(\lambda _{15} v_{\eta } v_{\Phi }-2 M v_{\chi }\right) & 
  \frac{2\lambda _9 v_{\eta }v_{\rho }^{2}-\lambda _{15} v_{\eta }v_{\Phi }^2+2\sqrt{2}v_{\rho }v_{\chi } f +2 M v_{\chi } v_{\Phi }}{4 v_{\eta }} & 
  \frac{1}{2} \left(\lambda _9 v_{\eta } v_{\rho }+\sqrt{2} f v_{\chi }\right) \\
  
  \frac{1}{4} \lambda _{16} v_{\rho } v_{\Phi } & 
  \frac{1}{2} \left(\lambda _9 v_{\eta } v_{\rho }+\sqrt{2} f v_{\chi }\right) &
  \frac{1}{4} \left(2\lambda _9 v_{\eta }^2+\lambda _{16} v_{\Phi }^2+\frac{2 \sqrt{2} f v_{\eta } v_{\chi }}{v_{\rho }}\right) \\
\end{array}
\right).    
\]
Here, $\Phi^+$ also decouples from the other charged scalars and gains mass $m^2_{\Phi^+} \approx \frac{1}{4}\frac{M v_\text{ew} v_{\chi}}{ v_{\Phi}}$.

The sextet has a doubly charged scalar, $\Delta^{++}$, whose mass is
\begin{equation}
M^2_{\Delta^{++}}=-\lambda_{11}v_\Phi^2 -\frac{\lambda_{15}v_\eta^2}{2}+\lambda_{16}v_\rho^2 -\frac{\lambda_{17}v_\chi^2 }{2}+\frac{M v_\eta v_\chi}{2v_\Phi} \approx \frac{M v_{\eta} v_\chi}{2 v_\Phi}\,.
\label{m++}
\end{equation}
In the regime of energy we are assuming here, all the scalars that compose the sextet $S$  decouple from each other and from  the other scalars, and are degenerate in mass. For the study of the spectrum of scalars of these triplets in any regime of energy, we refer the reader to Ref.~\cite{Pinheiro:2022bcs}. We finish this appendix by observing that, for an illustrative benchmark in which $v_\Phi=1$ eV, $M=0.1$ keV, $v_\eta=10^2$ GeV and $v_{\chi^{\prime}}=10^4$ GeV, the scalars that compose the  sextet have mass around $7$ TeV.

\section{Left-handed neutrino annihilation via $Z^\prime$}
\label{appendix:A}
In this appendix we present the non-standard contributions to the cross section of left-handed neutrino annihilation. The amplitude of the annihilation process $\nu_L + \bar \nu_L  \leftrightarrow e^- + e^+ $ can be written as
\begin{equation}
    \mathcal{M} = \mathcal{M}_Z + \mathcal{M}_W + \mathcal{M}_{Z^\prime}\,,
\end{equation}
where $\mathcal{M}_Z$, $\mathcal{M}_W$, and $\mathcal{M}_{Z^\prime}$ represent the amplitude of $\nu_L$ annihilation via $Z$, $W$, and $Z^\prime$, respectively. We can write the squared amplitude summed over final spin states and averaged over initial spins as
\begin{equation}
    |\mathcal{M}|^2_{\nu_L + \bar \nu_L \leftrightarrow e^- + e^+ } = |\mathcal{M}|^2_\text{SM} + |\mathcal{M}|^2_\text{331},
\end{equation}
where $ |\mathcal{M}|^2_\text{SM}$ represents the SM contribution with $Z$ and $W$, and $|\mathcal{M}|^2_\text{331}$ is defined as
\begin{equation}
    |\mathcal{M}|^2_\text{331} \equiv \mathcal{M}_Z^\dagger  \mathcal{M}_{Z^\prime} + \mathcal{M}_{Z^\prime}^\dagger \mathcal{M}_Z + \mathcal{M}_W^\dagger  \mathcal{M}_{Z^\prime} + \mathcal{M}_{Z^\prime}^\dagger \mathcal{M}_W + |\mathcal{M}|^2_{Z^\prime} \,.
\end{equation}
In the massless limit, the non-standard squared amplitude is
 \begin{align}
    |\mathcal{M}|^2_\text{331} & =  \frac{g^4 \cos^2 2 \theta_W  \left( \cos 4 \theta_W  \left(s^2+2 s t+2 t^2\right)+s^2+2 s t-4 t^2 \cos 2 \theta_W +4 t^2\right)}{32 \cos^4 \theta_W (2 \cos 2 \theta_W  +1)^2 \left(m_{Z^\prime}^2-s\right)^2} \nonumber \\ 
   & +\frac{g^4 \cos 2 \theta_W  \left(\cos 4 \theta_W  \left(s^2+2 s t+2 t^2\right)+s^2+2 s t-4 t^2 \cos 2 \theta_W +4 t^2\right)}{16 \cos^4 \theta_W (2 \cos 2 \theta_W  +1) \left(m_Z^2-s\right) \left(m_{Z^\prime}^2 - s\right)} \nonumber\\
   & +  \frac{g^4 \cos^2 2 \theta_W   (s+t)^2}{4 \cos^2 \theta_W (2 \cos 2\theta_W  +1) \left(m_W^2-t\right) \left(m_{Z^\prime}^2-s\right)} \;.
\end{align}
Then, the unpolarized cross section for a process $1 \, + \, 2 \to 3 \, + \, 4$ in the massless limit is given by
\begin{equation}
    \sigma_{1 \, + \, 2 \to 3 \, + \, 4} = \frac{1}{64 \pi^2 s} \int |\mathcal{M}|^2_{1 \, + \, 2 \to 3 \, + \, 4} d \Omega \,,
\end{equation}
where $d\Omega = d\cos \theta d\phi = 2 \pi \, d\cos \theta$. Then we obtain 
\begin{align}
    \sigma_{\nu_L \bar \nu_L \to l \bar l}^{Z^\prime} & =  \frac{g^4 s  \cos^2 2 \theta_W  (\cos 4 \theta_W  -2 \cos 2 \theta_W+2) }{768 \pi \cos^4 \theta_W  (2 \cos 2 \theta_W +1)^2 \left(m_{Z^\prime}^2-s\right)^2} \\ 
    & +\frac{g^4 s \cos 2 \theta_W (\cos 4 \theta_W  -2 \cos 2 \theta_W+2) }{384 \pi  \cos^4 \theta_W (2 \cos 2 \theta_W +1) \left(m_{Z}^2-s\right) \left(m_{Z^\prime}^2- s\right)} \\ 
    & -\frac{g^4 \cos ^22\theta_W \sec ^2\theta_W \left(s \left(2 m_W^2+3 s\right)+2 \left(m_W^2+s\right)^2 \left(\ln \left(m_W^2\right)-\ln \left(m_W^2+s\right)\right)\right)}{128 \pi  s^2 (2 \cos 2\theta_W+1) \left(m_{Z^\prime}^2-s\right)}\,,
\end{align}
where $l$ represents the standard charged leptons. As a good approximation we consider $s$ small compared with the mediating gauge bosons, so we obtain:
\begin{align}
    \sigma_{\nu_L \bar \nu_L \to l \bar l}^{Z^\prime} & =  \frac{g^4 s  \cos^2 2 \theta_W  (\cos 4 \theta_W  -2 \cos 2 \theta_W+2) }{768 \pi \cos^4 \theta_W  (2 \cos 2 \theta_W +1)^2 m_{Z^\prime}^4} \nonumber \\ 
    & +\frac{g^4 s \cos 2 \theta_W (\cos 4 \theta_W  -2 \cos 2 \theta_W+2) }{384 \pi  \cos^4 \theta_W (2 \cos 2 \theta_W +1) m_{Z}^2 m_{Z^\prime}^2} \\ 
    & +\frac{g^4 \cos ^22\theta_W  \left(m_W^2 + 2 s\right)}{128 \pi \cos^2 \theta_W (2 \cos 2\theta_W+1) m_{Z^\prime}^2 m_W^2} \nonumber \,.
\end{align}

\section{Right-handed neutrino annihilation} \label{app:sterile_neutrino}
In this appendix, we present the dominant squared amplitudes used in the calculation of the right-handed neutrino contribution to the effective number of neutrino species. These amplitudes play a crucial role in the thermalization of right-handed neutrinos.

The possible final states of the $2\to2$ right-handed neutrino pair annihilation include: the new gauge bosons, quarks, charged leptons and left-handed neutrinos. Disregarding $t$-channel mediators, we focus on the expressions mediated by $s$-channel $Z^\prime$ exchange, in the approximation $ m_{Z^\prime}^2 \gg s $.
These expressions are given by:
\begin{align}
	&|\mathcal{M}|^2_{\nu_R \bar \nu_R \to \nu_L \bar \nu_L}  =  \frac{g^4 t^2 \cos^2(2 \theta_W)}{4 m_{Z^\prime}^4 (2 \cos (2 \theta_W)  +1)^2} \\
& \label{nuRintochargedleptons} 	|\mathcal{M}|^2_{\nu_R \bar \nu_R \to l \bar l}  = \frac{g^4 \left(\cos(4 \theta_W)   \left(s^2+2 s t+2 t^2\right)+3 s^2-4 \cos (2 \theta_W)  (s+t)^2+6 s t+4 t^2\right)}{8 m_{Z^\prime}^4 (2 \cos(2 \theta_W)  +1)^2}\\ 
&	|\mathcal{M}|^2_{\nu_R \bar \nu_R \to q_{1\,,2} \bar q_{1\,,2}}  =   \frac{g^4 t^2}{36 m_{Z^\prime}^4} \,, \\ 
&	|\mathcal{M}|^2_{\nu_R \bar \nu_R \to q_{3} \bar q_{3}}  = \frac{g^4 t^2 \cos^4(\theta_W )}{m_{Z^\prime}^4 (2 \cos(2 \theta_W)  +1)^2}\,,
\end{align}
where $q_{1,2,3}$ represent the first, second, and third quark generations, respectively. Consequently, the relevant cross sections are \begin{align}
	&\sigma_{\nu_R \bar \nu_R \to \nu_L \bar \nu_L}  = \frac{g^4 s \cos ^2(2 \theta_W ) }{192 \pi  (2 \cos(2 \theta_W)  +1)^2 m_{Z^\prime}^4},\\ 
	&\sigma_{\nu_R \bar \nu_R \to l \bar l}  =  \frac{g^4 s ( \cos (4 \theta_W)  -2 \cos(2 \theta_W) + 2)}{192 \pi  (2 \cos(2 \theta_W)  +1)^2  m_{Z^\prime}^4},\\ 
	&\sigma_{\nu_R \bar \nu_R \to q_d \bar q_d}  = \frac{g^4 s}{1728 \pi   m_{Z^\prime}^4}, \\
	&\sigma_{\nu_R \bar \nu_R \to q_u \bar q_u}  = \frac{g^4 s  \cos ^4\theta_W }{48 \pi  (2 \cos  (2 \theta_W)  +1)^2 m_{Z^\prime}^4}\,.
\end{align}

The analysis presented in \autoref{sec:right_handed} excludes processes mediated by \( W^\prime \) and \( U^0 \), specifically 
\begin{equation}
\bar{\nu}_{i_R} + \nu_{i_R} \rightarrow W^\prime \rightarrow \bar{l}_i + l_i
\quad \text{and} \quad
\bar{\nu}_{i_R} + \nu_{i_R} \rightarrow U^0 \rightarrow \bar{\nu}_{j_L} + \nu_{j_L},
\end{equation}
where $i, j = 1, 2, 3 $ label the fermion generations, and $ l_i $ denotes the charged lepton flavors $ e,\, \mu,\, \tau $. We briefly examine the impact of these contributions below. The diagram mediated by $ W^\prime $ modifies \autoref{nuRintochargedleptons} into
\begin{align}
|\mathcal{M}|^2_{\nu_{i_R} \bar \nu_{i_R} \to l_j \bar l_j}&=
\frac{g^4 \left(\cos(4 \theta_W)   \left(s^2+2 s t+2 t^2\right)+3 s^2-4 \cos (2 \theta_W)  (s+t)^2+6 s t+4 t^2\right)}{8 (m_{Z^\prime}^2-s)^2 (2 \cos(2 \theta_W)  +1)^2}
\nonumber\\ &+\frac{g^4 t^2 \cos (2 \theta )}{2(2 \cos (2 \theta )+1) \left(m_{Z^\prime}^2-s\right) \left(m_{W^\prime}^2+s+t\right)}\delta_{ij} \nonumber\\ &+\frac{g^4 t^2}{4\left(m_{W^\prime}^2+s+t\right)^2} \delta_{ij} \label{tchannelamplitude}
\end{align}

\begin{figure}[!h]
    \centering
    \includegraphics[width=0.7\linewidth]{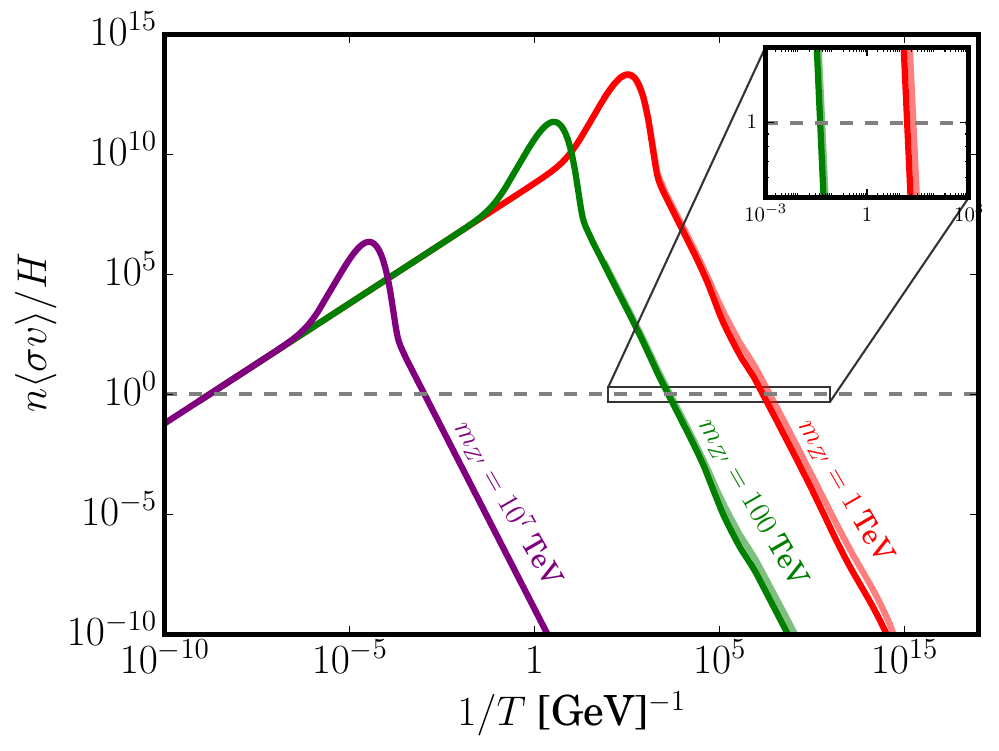}
    \caption{Ratio between annihilation rate of $\nu_R$  and the Hubble rate as a function of the inverse temperature for different masses of $Z^\prime$, indicated by the different colors. The dashed gray line represents $n\langle \sigma v \rangle /H =1$. The continuous curves represents the $Z^\prime$ contribution, \autoref{nuRintochargedleptons}, while the lightest curves include $W^\prime$ mediated $t$-channel annihilation, \autoref{tchannelamplitude}. }
    \label{rate_tchannel}
\end{figure}

In order to understand the contribution of $W^\prime$ to $\nu_R$  thermalization, we computed the interaction rate divided by the Hubble parameter, as shown in \autoref{rate_tchannel}. The solid curves represent the interaction rate mediated solely by $Z^\prime$, \autoref{nuRintochargedleptons}, while the lightest curves include the additional contributions from
$W^\prime$-mediated processes. As can be seen, the effect of $W^\prime$ is negligible in the regime of high  $Z^\prime$ masses, above $1$ TeV, which is the parameter space of interest in this work\footnote{It is worth emphasizing that the masses of 
$W^\prime$ and  $U^0$ can be expressed in terms of the $Z^\prime$  mass as $m_{W^\prime} \approx m_{U^0} \approx 0.72 m_{Z^\prime}$.}. Regarding the processes mediated by  $U^0$, their contribution is even more suppressed than that of $W^\prime$, as it depends on elements of the PMNS matrix, with $|U^\text{PMNS}_{ij}| \sim \mathcal{O}(10^{-1})$, entering either quadratically (via interference) or quartically (from pure $U^0$ exchange), thus providing an additional suppression factor. As expected, the inclusion of these t-channel diagrams results in only minor corrections, which do not qualitatively affect our conclusions.

\section{Sextet scalar contribution to $N_\text{eff}$}\label{ap.:D}

The scalars from the sextet also mediate interactions between right- and left-handed neutrinos and leptons, influencing the neutrino decoupling process and, consequently, the value of $N_\text{eff}$. However, as we demonstrate in this appendix, such contributions are negligible and can be safely neglected. In order to see this, we must, firstly,  open the Yukawa interaction in Eq. (\ref{yukawa2}), which give us the following set of interactions, 
\begin{align}
        - \mathcal{L}^Y \supset \frac{ G^\nu_{ab}}{\sqrt{2}} \left( \sqrt{2}\bar \nu_{a_L} \nu_{b_L}^C \Delta^0 + 2\bar \nu_{a_L} \nu_{b_R} \Phi^0 + \sqrt{2} \bar \nu_{a_R}^C \nu_{b_R} \sigma^0  +2\bar l_{a_L} \nu_{b_L}^c \Delta^+  + \right. \\ \left. +  2\bar l_{a_L} \nu_{b_R} \Phi^+ + \sqrt{2} \bar l_{a_L} l_{b_L}^C \Delta^{++}  \right) +  \text{H.c.} \,. \nonumber
\end{align}

Here, we focus on the interactions that could potentially affect neutrino decoupling. As previously mentioned, we neglect self-interactions, which is a good approximation as discussed in Ref.~\cite{Kawasaki:2000en}. Additionally, the lepton self-interaction mediated by $\Delta^{++}$ is not relevant for our purposes. Therefore, we consider only the following terms:
\begin{align}
        - \mathcal{L}^Y \supset \frac{ G^\nu_{ab}}{\sqrt{2}} \left(  2\bar \nu_{a_L} \nu_{b_R} \Phi^0  +2\bar l_{a_L} \nu_{b_L}^c \Delta^+  +  2\bar l_{a_L} \nu_{b_R} \Phi^+  \right) +  \text{H.c.} \,.
\end{align}

As previously noted, the scalars from the sextet contribute exclusively to scattering processes. According to the literature, such processes have a negligible effect on neutrino decoupling when annihilation channels are present. Furthermore, by definition, scattering processes yield $\delta n / \delta t = 0$ \cite{EscuderoAbenza:2020cmq}. Nevertheless, to illustrate their subdominant role compared to annihilation, we consider the scattering process $\bar{\nu}_L + \nu_R \leftrightarrow \bar{\nu}_L + \nu_R$ mediated by $\Phi^0$. It is important to note that $\Phi^0$ can be decomposed into its CP-even and CP-odd components. As shown in \autoref{appendix:A1}, these components decouple from the remaining scalars and share the same mass term, given by $m_{H^\prime}^2 = m_{A^\prime}^2 = \frac{1}{4} \frac{M v_\text{ew} v_\chi}{v_\Phi}$. After integrating out the mediator, we obtain:
\begin{equation}
    - \mathcal{L}^Y_\text{eff} \supset \frac{4 (G^\nu_{ab})^2}{2 m_{\Phi^0}^2} \, \bar \nu_{a_L} \nu_{b_R} \, \bar \nu_{a_L} \nu_{b_R}
\end{equation}

where $ m_{\Phi^0}^2 = m_{H^\prime}^2 $, and $ G^\nu_{ab} \sim \mathcal{O}(10^{-3} - 10^{-2}) $, depending on the neutrino mass ordering (normal or inverted). For the benchmark values $ v_{\chi^\prime} = 10~\text{TeV} $, $ v_\Phi = 1~\text{eV} $, and $ M = 0.1~\text{keV} $, we find \( m_{\Phi^0} \sim 7~\text{TeV} \), leading to an effective coupling of order $ \mathcal{O}(10^{-12})~\text{GeV}^{-2} $.

To compare this with the annihilation process $\bar{\nu}_L + \nu_L \leftrightarrow \bar{\nu}_R + \nu_R $ mediated by $Z^\prime$, we consider the effective interaction:

\begin{equation}
    - \mathcal{L}^{Z^\prime}_\text{eff} \supset \frac{g^2}{2 m_{Z^\prime}^2} \frac{1 - 2 s_W^2}{3 - 4 s_W^2} \, \bar \nu_{L} \nu_{L} \, \bar \nu_{R} \nu_{R}
\end{equation}

This yields an effective coupling of approximately  $\mathcal{O}(10^{-5})~\text{GeV}^{-2} $, which is about $ 10^{7} $ times larger than the contribution from the sextet.

Based on the arguments presented above, we can safely neglect the contribution of these scalars to neutrino decoupling. Their impact is strongly suppressed compared to the dominant annihilation processes, and thus they do not significantly affect the value of $N_\text{eff}$.

\bibliographystyle{JHEPfixed}
\bibliography{references}

\end{document}